\documentclass[fleqn,10pt]{wlscirep}
\usepackage[utf8]{inputenc}
\usepackage[T1]{fontenc}
\usepackage{float}
\usepackage{chemformula} 
\usepackage{mathpazo}
\usepackage{lineno}
\usepackage[normalem]{ulem}

\usepackage{xcolor}
\usepackage{graphicx}
\usepackage{geometry}
\usepackage{relsize}
\usepackage{multicol}
\usepackage{amsmath}
\usepackage{nccmath}
\title{Observation of phonon Poiseuille flow in isotopically-purified graphite ribbons}

\author[1,$\dag$]{Xin Huang}
\author[1,$\dag$]{Yangyu Guo}
\author[1]{Yunhui Wu}
\author[1]{Satoru Masubuchi}
\author[2]{Kenji Watanabe}
\author[1,3]{Takashi Taniguchi}
\author[1]{Zhongwei Zhang}
\author[1,4]{Sebastian Volz}
\author[1]{Tomoki Machida}
\author[1,5,*]{Masahiro Nomura}
\affil[1]{Institute of Industrial Science, The University of Tokyo, Tokyo 153-8505, Japan}
\affil[2]{Research Center for Functional Materials,
National Institute for Materials Science, Tsukuba 305-0044,
Japan}
\affil[3]{International Center for Materials
Nanoarchitectonics, National Institute for Materials Science,
Tsukuba 305-0044, Japan}
\affil[4]{LIMMS, CNRS-IIS IRL 2820, The University of Tokyo, Tokyo 153-8505, Japan}
\affil[5]{Research Center for Advanced Science and Technology, The University of Tokyo, Tokyo 153-0041, Japan}
\affil[*]{corresponding author: Masahiro Nomura (nomura@iis.u-tokyo.ac.jp)}

\affil[$\dag$]{these authors contributed equally to this work}

\begin{abstract}

In recent times, the unique collective transport physics of phonon hydrodynamics motivates theoreticians and experimentalists to explore it in micro- and nanoscale and at elevated temperatures. Graphitic materials have been predicted to facilitate hydrodynamic heat transport with their intrinsically strong normal scattering. However, owing to the experimental difficulties and vague theoretical understanding, the observation of phonon Poiseuille flow in graphitic systems remains challenging. In this study, based on a microscale experimental platform and the pertinent occurrence criterion in anisotropic solids, we demonstrate the phonon Poiseuille flow in a 5~\textmu m-wide suspended graphite ribbon with purified $^{13}$C isotope concentration. Our observation is well supported by our theoretical model based on a kinetic theory with fully first-principles inputs. Thus, this study paves the way for deeper insight into phonon hydrodynamics and cutting-edge heat manipulating applications.     

\end{abstract}

\makeatletter
\renewcommand{\fnum@figure}{Fig. \thefigure}
\makeatother

\begin{document}

\maketitle

\thispagestyle{empty}

\noindent The classical Fourier's law well describes the diffusive phonon transport in macroscale materials at high temperatures, where the frequent Umklapp phonon-phonon scatterings damp the heat flux. Cooling or down-scaling of the systems invalidates the Fourier's law and gives rise to non-Fourier heat transport behaviors \cite{wilson2014anisotropic,guo2015phonon,nomura2018thermal,chen2021non}, such as coherent \cite{luckyanova2012coherent,ravichandran2014crossover,zen2014engineering,maire2017heat}, ballistic \cite{anufriev2017heat,lee2015ballistic,vakulov2020ballistic}, and hydrodynamic \cite{cepellotti2015phonon,lee2015hydrodynamic,huberman2019observation,jeong2021transient,martelli2018thermal} transport. In contrast to ballistic or coherent phonon transport dictated by the boundary and interface, hydrodynamic transport is governed by intrinsically momentum-conserving normal phonon-phonon scattering. The frequent normal processes lead to exceptionally collective behaviors of phonons similar to those of fluids, including second sound in transient-state \cite{huberman2019observation,jeong2021transient} and phonon Poiseuille flow in steady-state  \cite{ding2018phonon,martelli2018thermal}. The theoretical prediction and experimental observation of phonon hydrodynamics in solids are of vital significance for both the fundamentals of lattice dynamics due to its unusual physics and the potential applications in thermal management due to its excellent transport properties.

The second sound, named analogously to the first sound (pressure wave), denotes the temperature wave propagating in solid-state materials \cite{hardy1970phonon,beardo2021observation}. The phonon Poiseuille flow is similar to that of viscous fluids under the pressure gradient in a pipe. The Poiseuille flow of phonons results from the interplay between normal scattering and diffuse boundary scattering events in the structure with a finite width. Phonon momenta are transferred along the gradient of drift velocity from the sample center to the sides by normal processes and destroyed at the boundaries \cite{lee2015hydrodynamic,li2018role,liao2020nanoscale}, inducing a parabolic heat flux profile (Fig.~\ref{lab_fig1}a). The experimental detection of second sound in solids has a long history and has been widely reported owing to its direct wavy feature. The drifting second sound was first observed unambiguously in solid He$^4$ crystals \cite{ackerman1966second}, later in various other crystals \cite{jackson1970second,rogers1971transport,narayanamurti1972observation,pohl1976observation,hehlen1995observation,koreeda2007second} with heat-pulse and light-scattering methods at low temperatures, whereas the driftless counterpart has been detected very recently in Ge even at room temperature under a rapidly varying temperature field \cite{beardo2021observation}. However, owing to the difficulty in observation and lack of direct evidence, there are limited experimental reports on the phonon Poiseuille flow \cite{smontara1996phonon,martelli2018thermal,machida2018observation}. Furthermore, this is also partially caused by the ambiguous criterion to confirm the evidence of phonon Poiseuille flow, as to be shown in the present study.  

Graphitic materials, owing to their intensive normal scattering due to the strong anharmonicity, and the high density of states of the low-lying flexural (or bending) phonon modes, are considered as the suitable systems for demonstrating phonon hydrodynamics at elevated temperatures \cite{li2018role,lee2015hydrodynamic,liao2020nanoscale}. The second sound has been recently observed in highly oriented pyrolytic graphite (HOPG) using transient thermal measurement techniques at recording high temperatures \cite{huberman2019observation,jeong2021transient,ding2022observation}. Despite the numerous theoretical investigations of phonon Poiseuille flow in graphitic materials \cite{guo2017heat,ding2018phonon,li2018role,guo2021size}, the experimental observation remains challenging owing to its more stringent observation window condition compared to that of the second sound, as to be elucidated in this work. It requires a special temperature range to realise the dominance of normal scattering and well-designed suspended microstructures to establish the hydrodynamic phonon flow. The indication of phonon Poiseuille flow was reported in a recent experimental work on bulk-scale natural graphite samples \cite{machida2020phonon}. However, it still remains inconclusive due to the pending theoretical explanation of the anomalous thickness-dependent trend and the ambiguous criterion. Additionally, the isotope-phonon scattering, as a momentum-destroying process, has been predicted to play an indispensable role in suppressing the occurrence of phonon Poiseuille flow \cite{lee2015hydrodynamic,cepellotti2015phonon}. However, the impact of isotope content in graphitic samples on the phonon hydrodynamic phenomena remains experimentally unexplored.

In this work, we present an unambiguous experimental evidence of phonon Poiseuille flow in graphitic materials. We design and fabricate submicroscale-suspended graphite ribbons and measure the thermal conductivity using a non-contact microsecond-scale time-domain thermoreflectance (\textmu-TDTR) technique. In addition, we investigate hydrodynamic phonon transport in both the natural and isotopically-purified graphite samples in a wide temperature range of 10$-$200~K. Supported by our first-principles-based theoretical modelling, we uncover the impact of the anisotropic nature of graphite on the criterion of phonon Poiseuille flow, and the appreciable influence of isotope content on its occurrence. 

\section*{Results}
\subsection*{Samples and thermal conductivity measurement}

Our isotopically-purified graphite crystal was synthesised using the high-pressure/high-temperature (HPHT) technique \cite{taniguchi2001spontaneous,taniguchi2007synthesis}, and the isotopic abundance of $^{13}$C was measured to be 0.02\% using a time-of-flight secondary ion mass spectrometry (TOF-SIMS) (Supplementary Note 1 and Supplementary Fig. 1a). The $^{13}$C isotope concentration in the natural graphite crystal was 1.1\%. A Raman spectroscopy was employed to characterise the crystallinity of the sample. As seen in Supplementary Fig. 1b, the Raman spectra show two main peaks for both crystals: the G peak at $\sim$1561 cm$^{-1}$ and the 2D peak at $\sim$2710 cm$^{-1}$, representing the sp$^2$ bonding of carbon atoms and perfect crystallite of the samples \cite{reich2004raman, ferrari2006raman, eckmann2012probing}. The only difference between these two samples is the isotopic concentration of $^{13}$C. Both samples are treated under the same conditions in the entire fabrication and measurement process. 

\begin{figure}[H]
\centering
\includegraphics[width=0.8\textwidth]{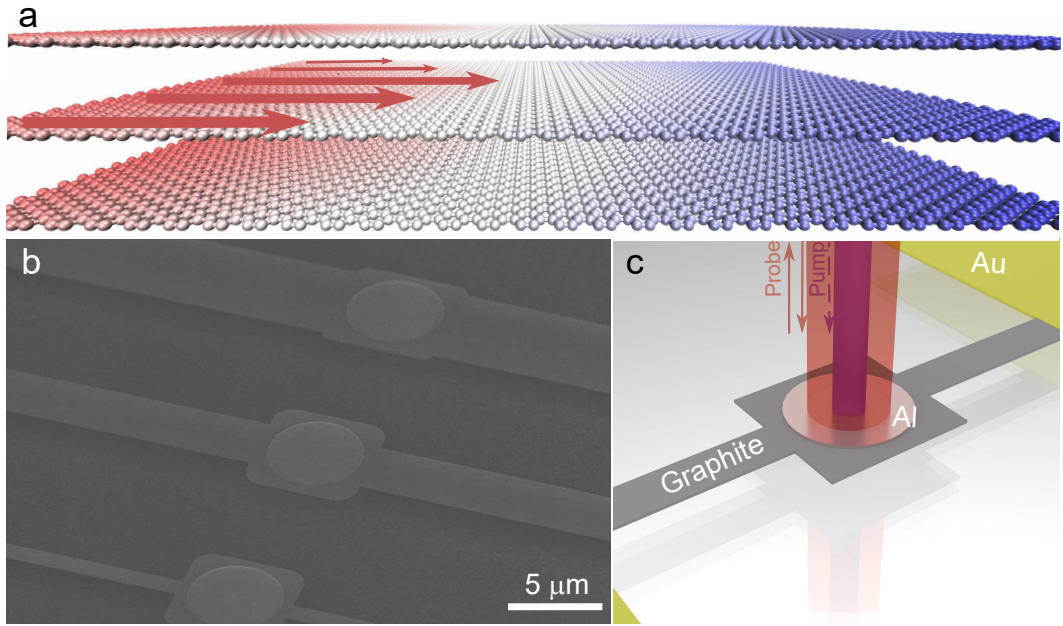}
\caption{\textbf{Isotopically-purified graphite ribbons and measurement method.} \textbf{a} Illustration of phonon Poiseuille flow in graphite ribbon. In the hydrodynamic regime, the heat flux (represented by the red arrows) manifests a parabolic profile maintained by the collective motion of phonons through a ribbon structure with a finite width. \textbf{b} SEM image of suspended graphite ribbons with various widths. \textbf{c} Schematic of the \textmu-TDTR measurement with the pump-probe method.}
\label{lab_fig1}
\end{figure}

Starting from a $\sim$50$\times$150~\textmu m$^2$ graphite flake, we first patterned and fabricated several suspended graphite ribbons connecting the graphite islands and the gold heat sinks from both sides (Fig.~\ref{lab_fig1}b). All the ribbons have the same length of 30 \textmu m, and various widths from 500 nm to 5 \textmu m. Next, we applied the pump-probe technique for investigating the in-plane phonon transport through the graphite ribbons with a circular aluminum transducer deposited on the central island, as depicted in Fig.~\ref{lab_fig1}c. Our \textmu-TDTR setup ensures the precise measurement of the steady-state thermal conduction in the ribbons at the microsecond-scale (see details in Methods). The thickness of the ribbons is identical to that of the initial flake, and it is estimated to be $\sim$65 nm from the cross-section view of the scanning electron microscope (SEM) image (Supplementary Fig. 1a). Other sample conditions, including the impurity concentration and surface roughness, are assumed to be consistent for all the adjacent ribbons from the same flake to justify the width-dependent investigation in the following content. 

\begin{figure}[H]
\centering
\includegraphics[width=0.65\textwidth]{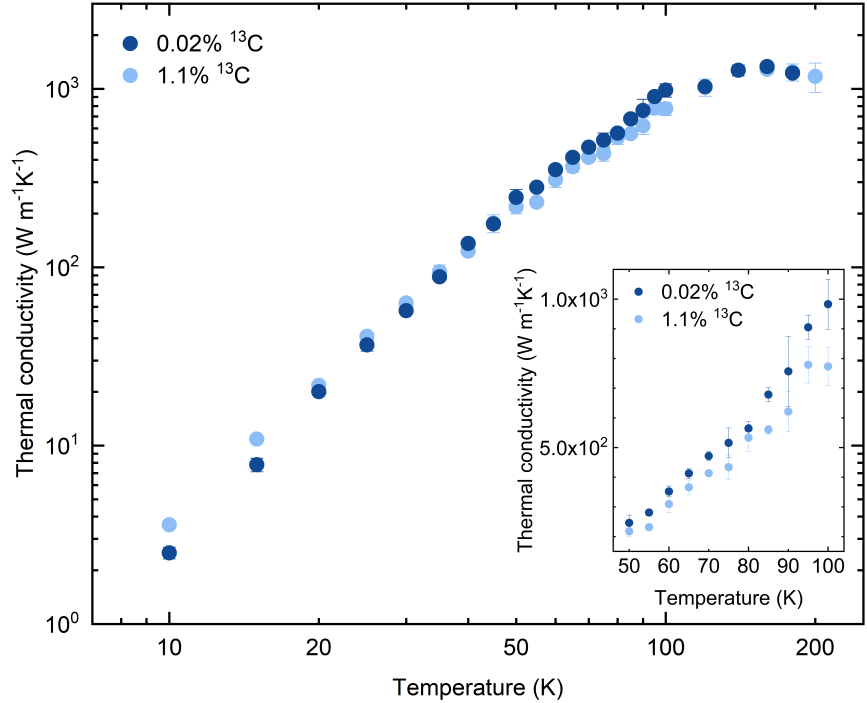}
\caption{\textbf{Temperature-dependent in-plane thermal conductivity of isotopically-purified and natural graphite ribbons with a designed width of 5~\textmu m.}  The dark blue and the light blue dots represent the results of isotopically-purified (0.02\% $^{13}$C) and natural (1.1\% $^{13}$C) graphite ribbons, respectively. Note that the actual width of the natural graphite ribbon is 1~\textmu m wider than that of the isotopically-purified one due to the deviation in fabrication, resulting in the minor flip of thermal conductivities at very low temperatures. Larger enhancement of thermal conductivity from isotope purification is expected when their widths are exactly the same. Inset: thermal conductivity in linear scale from 50 to 100~K.}
\label{lab_fig2}
\end{figure}

Figure~\ref{lab_fig2} shows the in-plane thermal conductivity of the isotopically-purified (0.02\% $^{13}$C) and natural (1.1\% $^{13}$C) graphite ribbons with the same designed width of 5~\textmu m. We measured the thermal conductivity of both the ribbons from 200 K down to 10 K, which sufficiently covers the temperature range of hydrodynamic window condition in graphite as reported in previous theoretical and experimental studies \cite{huberman2019observation, machida2020phonon, ding2018phonon}. With the decrease in temperature from 200~K, as the Umklapp phonon scattering becomes weaker, the thermal conductivities of both samples follow an increasing trend until they reach their peaks at around 150~K. The peak value of isotopically-purified sample is measured as $\sim$1330~Wm$^{-1}$K$^{-1}$, which is lower than that of bulk HOPG \cite{machida2020phonon} reported in a recent work owing to the strong size effect from structure down-scaling \cite{bae2013ballistic,xu2014length,fugallo2014thermal} (Supplementary Fig. 2). The phonon Umklapp process is further weakened below 100 K, where appreciable difference emerges between the thermal conductivities of the two samples with different $^{13}$C concentrations. The isotope scattering plays an important role in this regime, resulting in the enhanced thermal conductivity of isotopically-purified graphite ribbon compared to that of the natural one by 21\% at 100~K, which also qualitatively affects the occurrence of phonon Poiseuille flow as discussed in the following content. The isotopic effect remains prominent until the temperature goes down to 50 K, below which the thermal conductivities of the two samples become comparable again due to the gradual dominance of boundary scattering over isotope scattering. These trends of the thermal conductivities are similar in natural and isotopically-purified graphite ribbons with the widths of 1~\textmu m and 3~\textmu m, as shown in Supplementary Fig. 3. The effect of isotopic enrichment on the increase of thermal conductivity has also been observed in many other materials within the temperature range of 128$-$380~K, such as Si (10\%) \cite{inyushkin2004isotope}, GaN (15\%) \cite{zheng2019thermal}, boron phosphide (17\%) \cite{zheng2018high}, graphene (36\%) \cite{chen2012thermal}, and diamond (50\%) \cite{anthony1990thermal}.

\subsection*{Phonon Poiseuille flow}

In the hydrodynamic regime, the collective motion of thermal phonons due to the dominant intrinsic normal process demonstrates a Poiseuille flow of heat, as shown in Fig.~\ref{lab_fig1}a. In this regime, the probability of phonons losing momentum (due to resistive scattering) is notably reduced along their transport paths. In contrast, the phonon momentum is frequently destroyed in the ballistic regime due to the diffuse boundary-phonon scattering events. Therefore, a faster increase in thermal conductivity than the ballistic limit is considered as the indicator of the phonon Poiseuille flow \cite{lee2015hydrodynamic,cepellotti2015phonon,li2018role}. Quantitatively, a simple kinetic formula estimates the thermal conductivity as $\kappa$ $\sim$ $Cvl$, with $C$, $v$, and $l$ being the heat capacity, group velocity, and mean free path (MFP), respectively. The effective momentum-destroying MFP in the hydrodynamic transport can be obtained from the random walk theory as \cite{gurzhi1964thermal,gurzhi1968hydrodynamic}: $l\sim$ $W^2/l_{N}$, with $W$ as the sample width and $l_{N}$ as the MFP of normal process. One may feature the hydrodynamic thermal transport in principle from the temperature-dependence of $l$ associated with the strength of the normal process, as detected in some crystals \cite{kopylov1974investigation,machida2018observation}. The MFP is instead limited by the sample width ($l\sim$ $W$) in the ballistic limit: thus, $\kappa$ $\sim$ $CvW$. For most common three-dimensional (3D) materials, in the hydrodynamic regime, meaning that at very low temperatures, the heat capacity follows the well-known Debye $T^3$ law, and the group velocity is approximately the speed of sound as a constant. Therefore, as the temperature increases, with the enhancement of normal scattering ($l_{N}$ decreases), the thermal conductivity also boosts more rapidly than the ballistic limit ($T^3$). As a result, the increase in thermal conductivity, with a temperature-dependent exponent larger than $3$, has been adopted as a criterion to confirm the hydrodynamic phonon flow in several 3D crystals, such as SrTiO$_{3}$ ($T^{\sim 3.5}$) \cite{martelli2018thermal}, Bi ($T^{3.5}$) \cite{kopylov1974investigation}, and He$^{4}$ ($T^8$) \cite{mezhov1966measurement}, within the temperature ranges of 6$-$13~K, 1.5$-$2.4~K, and 0.7$-$0.9~K, respectively. For a two-dimensional (2D) system like graphene, the ballistic thermal conductance (or conductivity) follows $T^{1.68}$ \cite{bae2013ballistic}, and a faster increasing trend of thermal conductivity over $T^{1.68}$ is used to indicate the phonon Poiseuille flow, and it has been obtained by varying the width of graphene ribbon in a previous theoretical study \cite{li2018role}. 
 
However, the situation becomes quite different for graphite, where single graphene layers are bonded through weak van der Waals force. The anisotropic nature of graphite makes the temperature scaling of the thermal properties different from those of its 2D counterpart (graphene) and other isotropic 3D materials. The measured heat capacity of graphite deviates from the Debye law and shows a smaller power dependence of $T^{2.5}$ with an exponent between those of 2D and 3D systems \cite{alexander1980low}. In a recent experimental report with natural HOPG sample, an increase in the ratio of thermal conductivity over $T^{2.5}$ or over heat capacity with increasing temperature has been adopted to indicate the phonon hydrodynamics \cite{machida2020phonon}. In the following part, we will illustrate that the ballistic limit (or ballistic thermal conductance ($G_\textup{ballistic}$), equivalently) of graphite shows a different temperature dependence from that of heat capacity, i.e., different from $T^{2.5}$. Therefore, a faster increase of $\kappa$ than $G_\textup{ballistic}$ with increasing temperature is expected for the presence of phonon Poiseuille flow in graphite. Thus, we will demonstrate unambiguously the evidence of phonon Poiseuille flow in isotopically-purified graphite ribbons with sufficiently large widths.

\begin{figure}[H]
\centering
\includegraphics[width=1\textwidth]{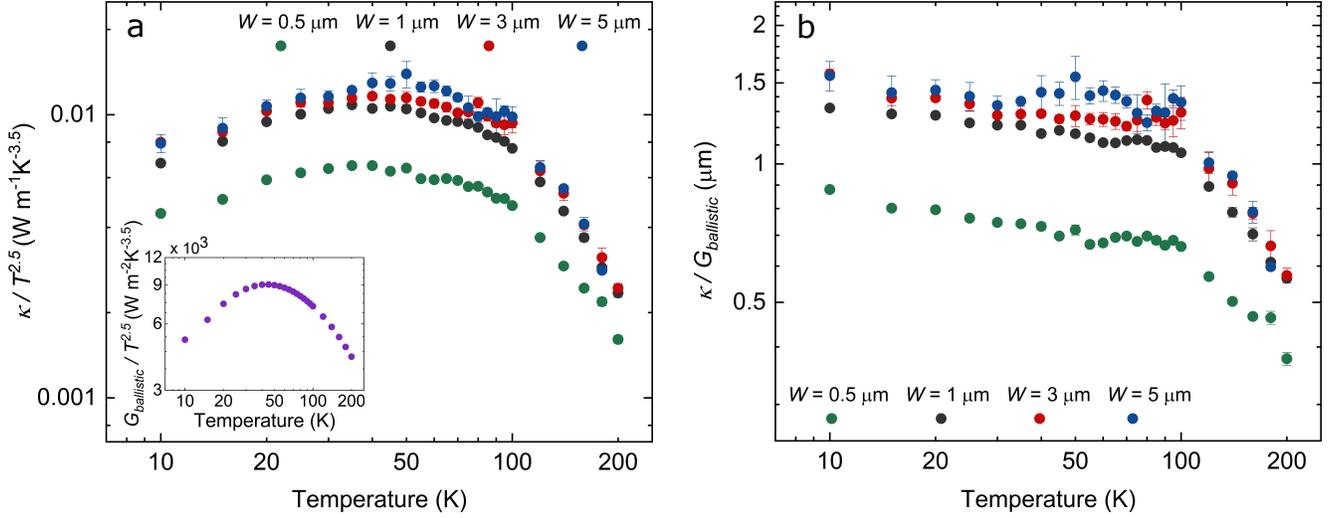}
\caption{\textbf{The criterion and evidence of phonon Poiseuille flow in isotopically-purified graphite ribbons.} \textbf{a} The usual criterion, namely the ratio of thermal conductivity ($\kappa$) over $T^{2.5}$ as a function of temperature ($T$) for graphite ribbons with various widths ($W$). Inset: temperature-dependence of ballistic thermal conductance ($G_\textup{ballistic}$) over $T^{2.5}$. \textbf{b} The present criterion, namely the ratio of thermal conductivity over $G_\textup{ballistic}$ as a function of temperature. The thermal conductivities of the isotopically-purified graphite ribbons are shown here.}
\label{lab_fig3}
\end{figure}

The results of $\kappa$/$T^{2.5}$ and $\kappa$/$G_\textup{ballistic}$ as a function of temperature are given in Fig.~\ref{lab_fig3}a and Fig.~\ref{lab_fig3}b, respectively, for our isotopically-purified graphite ribbons. The ballistic thermal conductance for graphite is calculated by the first-principles method (see details in Methods) as follow: 
\begin{ceqn}
\begin{equation}
 G_{ballistic} = \displaystyle\sum_{p}\int\limits v(\textup{\textbf{k}})\hbar \omega(\textup{\textbf{k}})\frac{\partial f^{eq}}{\partial T}\frac{d\textup{\textbf{k}}}{(2\pi)^{3}},\label{eqn1}
\end{equation}
\end{ceqn}
where $p$ represents phonon polarization, and $v(\textup{\textbf{k}})$, $\textbf{k}$, $\hbar$, $\omega(\textup{\textbf{k}})$, and $f^{eq}$ are the group velocity, wave vector, reduced Planck's constant, frequency, and Bose-Einstein equilibrium phonon distribution, respectively. Note that we also calculate the ballistic thermal conductance based on an empirical atomic interaction potential, which has minor influence on its temperature dependence at low temperature and on the conclusion in the present work. More detailed discussions are given in Supplementary Note 2 and Supplementary Fig. 4. As shown in Fig.~\ref{lab_fig3}a, $\kappa$/$T^{2.5}$ increases with temperature from 10 to $\sim$ 50~K for all four isotopically-purified graphite ribbons with different widths, including in the case of 500~nm-wide ribbon, where the heat transport should lie within the ballistic regime. This could be explained by the faster increase of the $G_\textup{ballistic}$ than $T^{2.5}$ in the same temperature range, as illustrated in the inset of Fig.~\ref{lab_fig3}a. In other words, a faster increase of $\kappa$ than $T^{2.5}$ or the heat capacity may not indicate the occurrence of hydrodynamic phonon flow definitely. Thus, a more relevant criterion to demonstrate phonon Poiseuille flow in graphite would be the temperature-dependent trend of $\kappa$/$G_\textup{ballistic}$, as shown in Fig.~\ref{lab_fig3}b. In the narrowest graphite ribbon with a width of 500~nm, $\kappa$/$G_\textup{ballistic}$ continuously decreases with increasing temperature, as a sign of the transition from ballistic to diffusive transport. In this case, the sample width is too narrow for enough normal scatterings to occur. The trend is similar in the graphite ribbon with a width of 1~\textmu m, although with a slightly smaller dropping slope below $\sim$ 80~K. However, in the graphite ribbon with a larger width of 3~\textmu m, an apparent flattening of $\kappa$/$G_\textup{ballistic}$ is observed from 30~K, where the normal scattering starts to play an increasing role to cancel and prevail the effect of momentum-destroying phonon scatterings. When the width of the graphite ribbon is sufficiently large as 5~\textmu m, the normal scattering becomes frequent while the resistive (Umklapp and isotope) scattering is still scarce, and $\kappa$/$G_\textup{ballistic}$ starts to increase with increasing temperature from 30 to $\sim$ 60~K. This super-ballistic scaling of thermal conductivity with temperature is a clear evidence of phonon Poiseuille flow. At higher temperatures ($>$ 100~K), $\kappa$/$G_\textup{ballistic}$ shows a dramatic decrease with increasing temperature in the graphite ribbons with all the widths due to the increasing rate of Umklapp scattering.

\begin{figure}[H]
\centering
\includegraphics[width=1\textwidth]{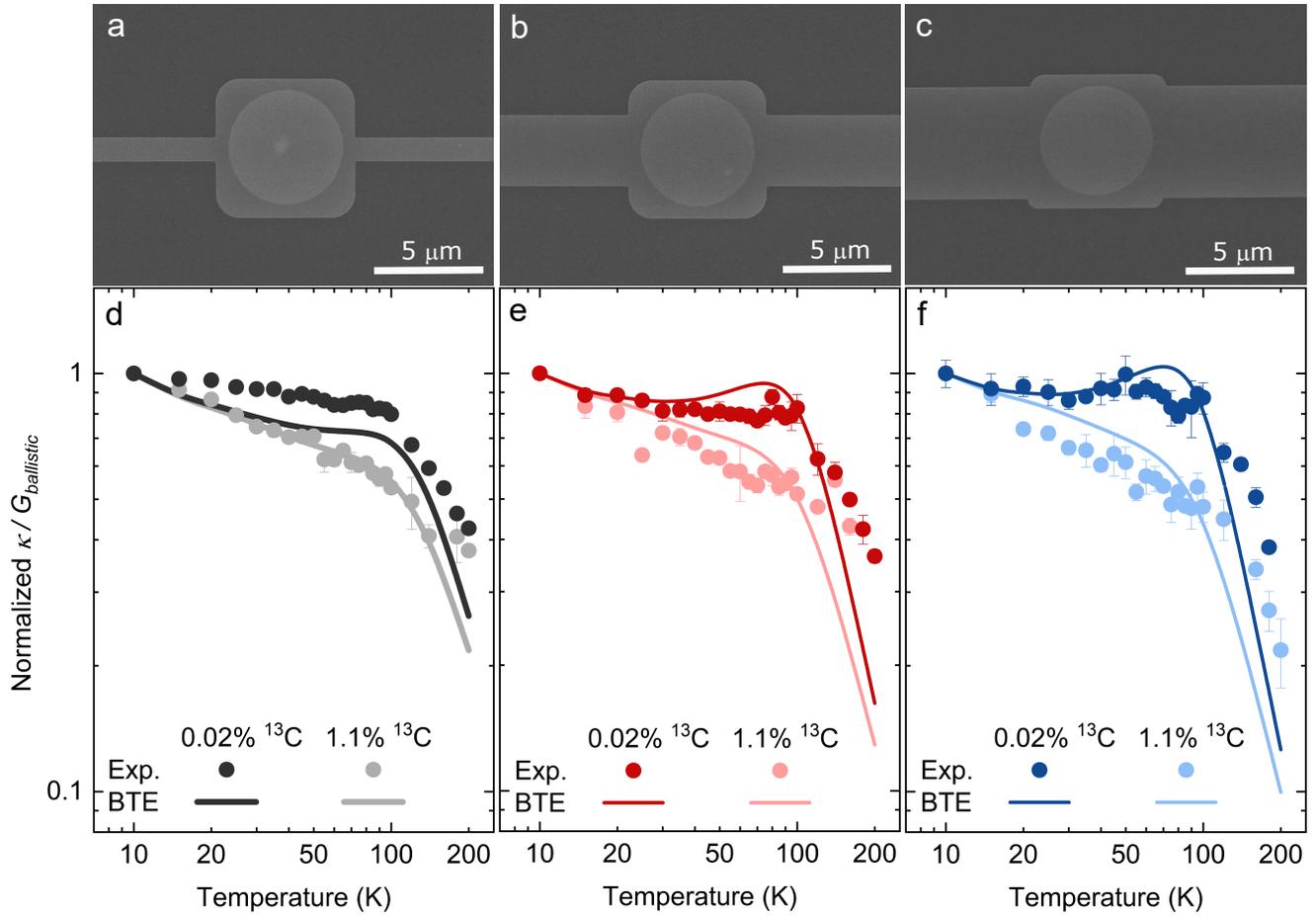}
\caption{\textbf{Isotope effect on phonon Poiseuille flow in graphite ribbons.} \textbf{a,b,c} SEM images of suspended isotopically-purified graphite ribbons with the width of 1~\textmu m, 3~\textmu m and 5~\textmu m, respectively. \textbf{d,e,f} Thermal conductivity over ballistic thermal conductance ($\kappa$/$G_\textup{ballistic}$) normalised by its value at 10~K as a function of temperature corresponding to the three ribbons in (\textbf{a,b,c}). The black (gray), red (pink), and dark blue (light blue) dots represent the experimental data of isotopically-purified (natural) graphite ribbons. The lines with the corresponding colors denote the modelling results by BTE with first-principles inputs.}
\label{lab_fig4}
\end{figure}

To observe the phonon Poiseuille flow, the rate of normal scattering should be dominant over those of the resistive ones, such as Umklapp scattering and phonon-isotope scattering \cite{lee2015hydrodynamic,cepellotti2015phonon}. To this end, we also investigated the isotope effect on the phonon Poiseuille flow by comparing the results of isotopically-purified (0.02\% $^{13}$C) and natural (1.1\% $^{13}$C) graphite ribbons. We first examined the occurrence of phonon Poiseuille flow based on the temperature dependence of $\kappa$/$G_\textup{ballistic}$ in 1~\textmu m-wide graphite ribbons (Fig.~\ref{lab_fig4}a). As illustrated in Fig.~\ref{lab_fig4}d, $\kappa$/$G_\textup{ballistic}$ of both samples show a decreasing trend as the temperature increases due to the predominant diffuse phonon-boundary scattering in relatively narrow ribbons. A steeper decrease is found in the natural graphite sample due to the additional effect of the momentum-destroying isotope scattering of phonons. Enlargement of the ribbon width to 3~\textmu m (Fig.~\ref{lab_fig4}b) makes the difference of $\kappa$/$G_\textup{ballistic}$ between isotopically-purified and natural samples more pronounced, as depicted in Fig.~\ref{lab_fig4}e. This is caused by the larger space for the phonon-isotope scattering to occur. When the ribbon width is sufficiently large, namely the 5~\textmu m case (Fig.~\ref{lab_fig4}c), $\kappa$/$G_\textup{ballistic}$ shows a non-monotonous trend in the isotopically-purified sample, while it continuously decreases in the natural one with increasing temperature, as shown in Fig.~\ref{lab_fig4}f. It infers that the phonon Poiseuille flow is deteriorated by the resistive phonon-isotope scattering in natural graphite sample. Within the temperature window of phonon Poiseuille flow (i.e., 30$-$60~K), the slope of $\kappa$/$G_\textup{ballistic}$ for isotopically-purified graphite ribbons evolves from a decreasing (1~\textmu m) to a flattening (3~\textmu m), and eventually an increasing (5~\textmu m) trend with the rise of temperature, indicating a clear transition from ballistic regime to hydrodynamic regime similar to the observation in a recent theoretical work \cite{ding2018phonon}. While the slope of $\kappa$/$G_\textup{ballistic}$ for the natural counterparts oppositely drops faster with the widening of the ribbon (as detailed in Supplementary Fig. 5). As temperature further increases beyond the hydrodynamic window, $\kappa$/$G_\textup{ballistic}$ would decrease due to the increasing Umklapp scattering rate. For the case of 5~\textmu m-wide isotopically-purified ribbon shown in Fig.~\ref{lab_fig4}f, the destruction of phonon Poiseuille flow occurs at the temperature over 60~K in our experimental result, which is slightly earlier than the drop of $\kappa$/$G_\textup{ballistic}$ at 70~K in our calculation. It might be caused by the unknown defect or contamination during sample fabrication which introduces additional resistive scattering and shortens the hydrodynamic temperature window to 30$-$60~K, or uncertainty in thermal conductivity measurement beyond 60~K. Besides, the aforementioned tendencies of experimental data are generally consistent with our theoretical modelling results based on a direct solution of phonon Boltzmann transport equation (BTE) with full first-principles inputs (see details in Methods).

\section*{Discussion}

The steady-state phonon hydrodynamic phenomenon, i.e., phonon Poiseuille flow, appears only when a strict condition is satisfied. It requires normal scattering to be more sufficient than boundary scattering, which is further more substantial than other resistive phonon scattering events, such as Umklapp and isotope scatterings. In nanoscale structures, the frequent interaction between phonons and the structure edges brings heat conduction to the ballistic regime. In much larger and isotopically-impure samples, the resistive scattering deteriorates the hydrodynamic phonon flow. Thus, the Poiseuille flow of phonons can be well-established in the graphite sample with a purified isotope concentration and a width in between the MFPs of normal and resistive scatterings \cite{guyer1966thermal} ($l_{N} \ll W, ~l_{R}l_{N}\gg W^{2}$). A recent theoretical work predicted that the width window condition of phonon hydrodynamics is approximately 2$-$20~\textmu m below 90~K in graphite ribbons with 0.1\% isotope content \cite{ding2018phonon}. Our observation of phonon Poiseuille flow in a 5~\textmu m-wide isotopically-purified graphite ribbon, therefore, confirms this theoretical prediction. We also provide a detailed quantitative demonstration of why the hydrodynamic window condition is satisfied only in the isotopically-purified graphite ribbon in Supplementary Note 2 and Supplementary Fig. 6. On the other hand, it explicitly demonstrates a more stringent condition for the observation of phonon Poiseuille flow than that of the second sound, which has been observed instead in natural graphite recently\cite{huberman2019observation,jeong2021transient,ding2022observation} (detailed explanation in Supplementary Note 3).

We have shown that the temperature dependence of $\kappa$/$G_\textup{ballistic}$ (or equivalently $\kappa$/$\kappa_\textup{ballistic}$) is a more relevant criterion to confirm the phonon Poiseuille flow. In most 3D materials close to isotropic structures, the ballistic thermal conductance and heat capacity follow the same temperature power law at low temperatures due to the linear dispersion relation of acoustic phonons \cite{chen2005,kittel2004,li2018role}. The different temperature scalings between $G_\textup{ballistic}$ and heat capacity ($T^{2.5}$) of graphite, shown in the inset of Fig.~\ref{lab_fig3}a, is mainly attributed to the anisotropic nature and special phonon dispersion of graphite. The hydrodynamic phonon transport is mainly contributed by the bending acoustic (BA) modes in graphite \cite{ding2018phonon,guo2021size}, the group velocity of which increases with frequency due to the quadratic dispersion curve \cite{alofi2014theory,lee2015hydrodynamic}. As expressed in equation (\ref{eqn1}), $G_\textup{ballistic}$ is determined by both the heat capacity term and the group velocity term ($v(\textup{\textbf{k}})$). Therefore, with an increase in temperature in the hydrodynamic window and more populated higher-frequency BA phonons, $G_\textup{ballistic}$ of graphite boosts faster than the heat capacity ($T^{2.5}$). As a result, the increase of $\kappa$/$T^{2.5}$ with temperature from few Kelvins to $\sim$ 25~K in the recent experimental report \cite{machida2020phonon} is most probably contributed by the behaviour of the ballistic thermal conductance. 

In addition, as a reference, we examined the isotope effect on the observation of phonon Poiseuille flow based on the present criterion ($\kappa$/$G_\textup{ballistic}$) in silicon samples with natural and purified $^{28}$Si isotope concentrations \cite{inyushkin2004isotope} (Supplementary Fig. 7). As temperature increases, $\kappa$/$G_\textup{ballistic}$ of both natural and purified silicon samples drops monotonically, indicating the absence of phonon Poiseuille flow even in the isotopically-purified silicon sample. This is explained by the well-known insufficient normal scattering in silicon to satisfy the hydrodynamic window condition.

According to the different temperature-dependent behaviors of $\kappa$/$G_\textup{ballistic}$ in our isotopically-purified graphite ribbons with various widths, we observe the transition from the ballistic to the hydrodynamic thermal transport when the ribbon width increases from 500~nm to 5~\textmu m. Besides, another important aspect to evidence the phonon hydrodynamic flow is the super-ballistic width dependence of thermal conductivity, as clearly indicated by the effective mean free path $l\sim$ $W^2/l_{N}$. The suspended microstructure system built up in this work provides a good platform for further experiments to investigate the extraordinary super-linear width dependence of thermal conduction in the hydrodynamic regime or the phonon Knudsen minimum phenomenon \cite{guo2017heat,ding2018phonon,li2019crossover,guo2021size}.

In summary, we have developed an integrated experimental platform to investigate the steady-state phonon hydrodynamics in suspended graphite submicron structures. In light of a more relevant criterion due to the anisotropic nature of graphite, we observed prominent Poiseuille flow of phonons in an isotopically-purified graphite ribbon with a width of 5~\textmu m up to 60~K, which is much elevated compared to the temperature range of previous observations in other solid-state materials \cite{mezhov1966measurement,kopylov1974investigation,bausch1972thermal,machida2018observation}. The phonon Poiseuille flow is prone to be destroyed by the resistive phonon-isotope scattering in graphite ribbons with natural abundance of carbon isotope. Our joint theoretical and experimental study on phonon hydrodynamics in graphitic materials thus deepens the understanding of the collective physics of phonons in anisotropic solids. The experimental platform will also open innovative possibilities for tuning and manipulation of phonon hydrodynamics, as well as its application in thermal management of the modern micro- and nanoelectronics. 

\section*{Methods}

\subsection*{Sample preparation}

Sample fabrication began with mechanical exfoliation of graphite to obtain graphite flakes, which were then transferred onto a SiO$_{2}$ (2.4~\textmu m)/Si substrate right after O$_{2}$ plasma surface treatment. The typical size of a flake was $\sim$50$\times$150~\textmu m$^2$, and the flake thickness is approximately 65~nm (as shown in Supplementary Fig. 1a), measured by SEM. Next, we applied electron-beam lithography (EBL) for patterning ribbon structures with a fixed length of 40~\textmu m but varying widths. Furthermore, a 6$\times$6~\textmu m$^2$ graphite island was patterned in the center of the ribbons on the same flake. We also used an electron-beam physical vapor deposition (EBPVD) to deposit 100 nm-thick aluminum on top of the graphite ribbons as masks for O$_{2}$ plasma etching. After exposing samples in a reactive ion etching chamber with an O$_{2}$ plasma source, we removed the rest of the graphite around the ribbon structures and released the aluminum masks to acquire the desired graphite ribbons. Moreover, we used laser lithography and another EBPVD to fabricate two 250$\times$400~\textmu m$^2$ gold pads used as hydrofluoric acid (HF) stoppers, to cover all the ribbons by 10~\textmu m-long from both sides. Following this, 70~nm-thick circular aluminum transducers with radius of 2.5~\textmu m were deposited on the central graphite island for TDTR measurement. HF vapor etching was used to remove SiO$_{2}$ from the entire surface, with only a portion of SiO$_{2}$ remaining underneath the gold pads to support the pads. Being attached and clamped by the gold pads from both sides, the ribbons were finally suspended for investigating phonon hydrodynamic transport, with a length of 30~\textmu m staying completely out-of-contact.

\subsection*{Thermal characterisation}

We employed a well-developed \textmu-TDTR setup to measure the thermal conductivity of our samples \cite{maire2014reduced}. We placed our samples in a vacuum cryostat with a pressure below 10$^{-5}$~Pa to avoid the convective heat loss. At 200~K, radiation heat loss from graphite ribbons was estimated as $\sim$1~nW, whereas the input power from the pump beam is $\sim$200~nW. At temperatures below 150~K, the loss through radiation is negligible \cite{pope2001description,maire2015thermal}. A liquid helium flow system enables to cool down the cryostat to 4~K. A cryogenic temperature controller from Oxford Instruments adjusted the temperature with a precision of 0.1~mK. 

In the \textmu-TDTR setup, two lasers were focused onto the center of the aluminum transducer located on the graphite island: one pulsed pump beam with a wavelength of 642~nm and one continuous probe beam with a wavelength of 785~nm. The pump beam with a 2~\textmu s pulse duration at 1~kHz repetition rate induced excitation in the transducer. The probe beam measured the reflectivity change from the base value generated by the pump beam, and a photoreceiver monitored its reflection with a maximum bandwidth of 200~MHz to detect the response every 5~ns. An oscilloscope captured the probe signal averaging 10$^{4}$ measurements into a thermal decay as a function of time, as shown in Supplementary Fig. 8. The fitting parameter, $\tau$, is an indicator of the inherent thermal properties of the measured sample. We measured each ribbon with one single-width three to five times and calculated its standard deviation as depicted by the error bars in figures containing the experimental data.

To extract the thermal conductivity of the graphite ribbons, we built up a numerical model identical to the actual experimental one using the finite element method (in COMSOL Multiphysics). In the model, the material parameters of metals (gold and aluminum) are assumed as their bulk values due to neligible size effects, whereas the specific heat of graphite is taken from the reported benchmark data \cite{nihira2003temperature}. The anisotropic nature of heat conduction in graphite is taken into account by setting the bulk out-of-plane thermal conductivity from the literature \cite{ho1972thermal} and leaving the in-plane thermal conductivity as the only fitting parameter. Thermal boundary conductance between metals and graphite is also regarded as a crucial parameter in the model to correctly reproduce the experimental measurement, as further explained in details in Supplementary Note 4 and Supplementary Fig. 9. By injecting a 2~\textmu s heat flux pulse with a Gaussian spatial distribution onto the aluminum transducer, we simulated the heat dissipation through the graphite ribbon structures to reproduce the decay times in the experiments by sweeping the values of in-plane thermal conductivity. The in-plane thermal conductivity of graphite ribbon is obtained when an optimal fitting was found between the decay times obtained by simulation and by experiment (Supplementary Fig. 8). 

\subsection*{Theoretical modeling}

We modelled the hydrodynamic heat transport through finite-size graphite ribbons by directly solving the phonon Boltzmann transport equation (BTE) with fully first-principles inputs using the methodology developed in our recent work \cite{guo2021size}. The required input of phonon properties (including phonon dispersion relations and scattering rates) are calculated in the open-source package SHENGBTE \cite{li2014shengbte}, with the harmonic and anharmonic (third-order) force constants obtained by the first-principles calculations implemented in the package QUANTUM ESPRESSO \cite{giannozzi2009quantum}. To accurately describe the inter-layer interaction in graphite, we adopted the most advanced van der Waals non-local density functional theory, with all the numerical details and validation of the first-principle calculations provided in our previous work \cite{guo2021size}. The first-principles phonon properties were also used in the calculation of the ballistic thermal conductance in equation (\ref{eqn1}). In the direct numerical solution of phonon BTE, we modelled the graphite ribbon with the same length, width, and carbon isotope concentration as those in the experimental measurement. Fully diffuse scattering of phonons at the transverse boundaries of the graphite ribbon were considered. This was reasonable as the edges of the ribbon are generally rough. We assumed negligible size effect along the cross-plane direction of graphite ribbon with the present thickness of 65~nm ($\sim$ 200 graphene mono-layers) as justified in previous studies \cite{lindsay2011flexural,ghosh2010dimensional}.

\section*{Data availability}
The data that support the findings of this study are available from the corresponding authors upon reasonable request.

\bibliography{Manuscript}

\begin{thebibliography}{10}
\urlstyle{rm}
\expandafter\ifx\csname url\endcsname\relax
  \def\url#1{\texttt{#1}}\fi
\expandafter\ifx\csname urlprefix\endcsname\relax\def\urlprefix{URL }\fi
\expandafter\ifx\csname doiprefix\endcsname\relax\def\doiprefix{DOI: }\fi
\providecommand{\bibinfo}[2]{#2}
\providecommand{\eprint}[2][]{\url{#2}}

\bibitem{wilson2014anisotropic}
\bibinfo{author}{Wilson, R.} \& \bibinfo{author}{Cahill, D.~G.}
\newblock \bibinfo{journal}{\bibinfo{title}{Anisotropic failure of fourier
  theory in time-domain thermoreflectance experiments}}.
\newblock {\emph{\JournalTitle{Nature Communications}}}
  \textbf{\bibinfo{volume}{5}}, \bibinfo{pages}{1--11} (\bibinfo{year}{2014}).

\bibitem{guo2015phonon}
\bibinfo{author}{Guo, Y.} \& \bibinfo{author}{Wang, M.}
\newblock \bibinfo{journal}{\bibinfo{title}{Phonon hydrodynamics and its
  applications in nanoscale heat transport}}.
\newblock {\emph{\JournalTitle{Physics Reports}}}
  \textbf{\bibinfo{volume}{595}}, \bibinfo{pages}{1--44}
  (\bibinfo{year}{2015}).

\bibitem{nomura2018thermal}
\bibinfo{author}{Nomura, M.}, \bibinfo{author}{Shiomi, J.},
  \bibinfo{author}{Shiga, T.} \& \bibinfo{author}{Anufriev, R.}
\newblock \bibinfo{journal}{\bibinfo{title}{Thermal phonon engineering by
  tailored nanostructures}}.
\newblock {\emph{\JournalTitle{Japanese Journal of Applied Physics}}}
  \textbf{\bibinfo{volume}{57}}, \bibinfo{pages}{080101}
  (\bibinfo{year}{2018}).

\bibitem{chen2021non}
\bibinfo{author}{Chen, G.}
\newblock \bibinfo{journal}{\bibinfo{title}{Non-fourier phonon heat conduction
  at the microscale and nanoscale}}.
\newblock {\emph{\JournalTitle{Nature Reviews Physics}}} \bibinfo{pages}{1--15}
  (\bibinfo{year}{2021}).

\bibitem{luckyanova2012coherent}
\bibinfo{author}{Luckyanova, M.~N.} \emph{et~al.}
\newblock \bibinfo{journal}{\bibinfo{title}{Coherent phonon heat conduction in
  superlattices}}.
\newblock {\emph{\JournalTitle{Science}}} \textbf{\bibinfo{volume}{338}},
  \bibinfo{pages}{936--939} (\bibinfo{year}{2012}).

\bibitem{ravichandran2014crossover}
\bibinfo{author}{Ravichandran, J.} \emph{et~al.}
\newblock \bibinfo{journal}{\bibinfo{title}{Crossover from incoherent to
  coherent phonon scattering in epitaxial oxide superlattices}}.
\newblock {\emph{\JournalTitle{Nature Materials}}}
  \textbf{\bibinfo{volume}{13}}, \bibinfo{pages}{168--172}
  (\bibinfo{year}{2014}).

\bibitem{zen2014engineering}
\bibinfo{author}{Zen, N.}, \bibinfo{author}{Puurtinen, T.~A.},
  \bibinfo{author}{Isotalo, T.~J.}, \bibinfo{author}{Chaudhuri, S.} \&
  \bibinfo{author}{Maasilta, I.~J.}
\newblock \bibinfo{journal}{\bibinfo{title}{Engineering thermal conductance
  using a two-dimensional phononic crystal}}.
\newblock {\emph{\JournalTitle{Nature Communications}}}
  \textbf{\bibinfo{volume}{5}}, \bibinfo{pages}{1--9} (\bibinfo{year}{2014}).

\bibitem{maire2017heat}
\bibinfo{author}{Maire, J.} \emph{et~al.}
\newblock \bibinfo{journal}{\bibinfo{title}{Heat conduction tuning by wave
  nature of phonons}}.
\newblock {\emph{\JournalTitle{Science Advances}}}
  \textbf{\bibinfo{volume}{3}}, \bibinfo{pages}{e1700027}
  (\bibinfo{year}{2017}).

\bibitem{anufriev2017heat}
\bibinfo{author}{Anufriev, R.}, \bibinfo{author}{Ramiere, A.},
  \bibinfo{author}{Maire, J.} \& \bibinfo{author}{Nomura, M.}
\newblock \bibinfo{journal}{\bibinfo{title}{Heat guiding and focusing using
  ballistic phonon transport in phononic nanostructures}}.
\newblock {\emph{\JournalTitle{Nature Communications}}}
  \textbf{\bibinfo{volume}{8}}, \bibinfo{pages}{1--8} (\bibinfo{year}{2017}).

\bibitem{lee2015ballistic}
\bibinfo{author}{Lee, J.}, \bibinfo{author}{Lim, J.} \& \bibinfo{author}{Yang,
  P.}
\newblock \bibinfo{journal}{\bibinfo{title}{Ballistic phonon transport in holey
  silicon}}.
\newblock {\emph{\JournalTitle{Nano Letters}}} \textbf{\bibinfo{volume}{15}},
  \bibinfo{pages}{3273--3279} (\bibinfo{year}{2015}).

\bibitem{vakulov2020ballistic}
\bibinfo{author}{Vakulov, D.} \emph{et~al.}
\newblock \bibinfo{journal}{\bibinfo{title}{Ballistic phonons in ultrathin
  nanowires}}.
\newblock {\emph{\JournalTitle{Nano Letters}}} \textbf{\bibinfo{volume}{20}},
  \bibinfo{pages}{2703--2709} (\bibinfo{year}{2020}).

\bibitem{cepellotti2015phonon}
\bibinfo{author}{Cepellotti, A.} \emph{et~al.}
\newblock \bibinfo{journal}{\bibinfo{title}{Phonon hydrodynamics in
  two-dimensional materials}}.
\newblock {\emph{\JournalTitle{Nature Communications}}}
  \textbf{\bibinfo{volume}{6}}, \bibinfo{pages}{1--7} (\bibinfo{year}{2015}).

\bibitem{lee2015hydrodynamic}
\bibinfo{author}{Lee, S.}, \bibinfo{author}{Broido, D.},
  \bibinfo{author}{Esfarjani, K.} \& \bibinfo{author}{Chen, G.}
\newblock \bibinfo{journal}{\bibinfo{title}{Hydrodynamic phonon transport in
  suspended graphene}}.
\newblock {\emph{\JournalTitle{Nature Communications}}}
  \textbf{\bibinfo{volume}{6}}, \bibinfo{pages}{1--10} (\bibinfo{year}{2015}).

\bibitem{huberman2019observation}
\bibinfo{author}{Huberman, S.} \emph{et~al.}
\newblock \bibinfo{journal}{\bibinfo{title}{Observation of second sound in
  graphite at temperatures above 100 \uppercase{K}}}.
\newblock {\emph{\JournalTitle{Science}}} \textbf{\bibinfo{volume}{364}},
  \bibinfo{pages}{375--379} (\bibinfo{year}{2019}).

\bibitem{jeong2021transient}
\bibinfo{author}{Jeong, J.}, \bibinfo{author}{Li, X.}, \bibinfo{author}{Lee,
  S.}, \bibinfo{author}{Shi, L.} \& \bibinfo{author}{Wang, Y.}
\newblock \bibinfo{journal}{\bibinfo{title}{Transient hydrodynamic lattice
  cooling by picosecond laser irradiation of graphite}}.
\newblock {\emph{\JournalTitle{Phys. Rev. Lett.}}}
  \textbf{\bibinfo{volume}{127}}, \bibinfo{pages}{085901}
  (\bibinfo{year}{2021}).

\bibitem{martelli2018thermal}
\bibinfo{author}{Martelli, V.}, \bibinfo{author}{Jim{\'e}nez, J.~L.},
  \bibinfo{author}{Continentino, M.}, \bibinfo{author}{Baggio-Saitovitch, E.}
  \& \bibinfo{author}{Behnia, K.}
\newblock \bibinfo{journal}{\bibinfo{title}{Thermal transport and phonon
  hydrodynamics in strontium titanate}}.
\newblock {\emph{\JournalTitle{Physical Review Letters}}}
  \textbf{\bibinfo{volume}{120}}, \bibinfo{pages}{125901}
  (\bibinfo{year}{2018}).

\bibitem{ding2018phonon}
\bibinfo{author}{Ding, Z.} \emph{et~al.}
\newblock \bibinfo{journal}{\bibinfo{title}{Phonon hydrodynamic heat conduction
  and knudsen minimum in graphite}}.
\newblock {\emph{\JournalTitle{Nano Letters}}} \textbf{\bibinfo{volume}{18}},
  \bibinfo{pages}{638--649} (\bibinfo{year}{2018}).

\bibitem{hardy1970phonon}
\bibinfo{author}{Hardy, R.~J.}
\newblock \bibinfo{journal}{\bibinfo{title}{Phonon boltzmann equation and
  second sound in solids}}.
\newblock {\emph{\JournalTitle{Physical Review B}}}
  \textbf{\bibinfo{volume}{2}}, \bibinfo{pages}{1193} (\bibinfo{year}{1970}).

\bibitem{beardo2021observation}
\bibinfo{author}{Beardo, A.} \emph{et~al.}
\newblock \bibinfo{journal}{\bibinfo{title}{Observation of second sound in a
  rapidly varying temperature field in \ch{Ge}}}.
\newblock {\emph{\JournalTitle{Science Advances}}}
  \textbf{\bibinfo{volume}{7}}, \bibinfo{pages}{eabg4677}
  (\bibinfo{year}{2021}).

\bibitem{li2018role}
\bibinfo{author}{Li, X.} \& \bibinfo{author}{Lee, S.}
\newblock \bibinfo{journal}{\bibinfo{title}{Role of hydrodynamic viscosity on
  phonon transport in suspended graphene}}.
\newblock {\emph{\JournalTitle{Physical Review B}}}
  \textbf{\bibinfo{volume}{97}}, \bibinfo{pages}{094309}
  (\bibinfo{year}{2018}).

\bibitem{liao2020nanoscale}
\bibinfo{editor}{Liao, B.} (ed.) \emph{\bibinfo{title}{Nanoscale Energy
  Transport}}.
\newblock 2053-2563 (\bibinfo{publisher}{IOP Publishing},
  \bibinfo{year}{2020}).

\bibitem{ackerman1966second}
\bibinfo{author}{Ackerman, C.~C.}, \bibinfo{author}{Bertman, B.},
  \bibinfo{author}{Fairbank, H.~A.} \& \bibinfo{author}{Guyer, R.~A.}
\newblock \bibinfo{journal}{\bibinfo{title}{Second sound in solid helium}}.
\newblock {\emph{\JournalTitle{Phys. Rev. Lett.}}}
  \textbf{\bibinfo{volume}{16}}, \bibinfo{pages}{789--791}
  (\bibinfo{year}{1966}).

\bibitem{jackson1970second}
\bibinfo{author}{Jackson, H.~E.}, \bibinfo{author}{Walker, C.~T.} \&
  \bibinfo{author}{McNelly, T.~F.}
\newblock \bibinfo{journal}{\bibinfo{title}{Second sound in \ch{NaF}}}.
\newblock {\emph{\JournalTitle{Physical Review Letters}}}
  \textbf{\bibinfo{volume}{25}}, \bibinfo{pages}{26} (\bibinfo{year}{1970}).

\bibitem{rogers1971transport}
\bibinfo{author}{Rogers, S.}
\newblock \bibinfo{journal}{\bibinfo{title}{Transport of heat and approach to
  second sound in some isotopically pure alkali-halide crystals}}.
\newblock {\emph{\JournalTitle{Physical Review B}}}
  \textbf{\bibinfo{volume}{3}}, \bibinfo{pages}{1440} (\bibinfo{year}{1971}).

\bibitem{narayanamurti1972observation}
\bibinfo{author}{Narayanamurti, V.} \& \bibinfo{author}{Dynes, R.}
\newblock \bibinfo{journal}{\bibinfo{title}{Observation of second sound in
  bismuth}}.
\newblock {\emph{\JournalTitle{Physical Review Letters}}}
  \textbf{\bibinfo{volume}{28}}, \bibinfo{pages}{1461} (\bibinfo{year}{1972}).

\bibitem{pohl1976observation}
\bibinfo{author}{Pohl, D.~W.} \& \bibinfo{author}{Irniger, V.}
\newblock \bibinfo{journal}{\bibinfo{title}{Observation of second sound in
  \ch{NaF} by means of light scattering}}.
\newblock {\emph{\JournalTitle{Physical Review Letters}}}
  \textbf{\bibinfo{volume}{36}}, \bibinfo{pages}{480} (\bibinfo{year}{1976}).

\bibitem{hehlen1995observation}
\bibinfo{author}{Hehlen, B.}, \bibinfo{author}{P{\'e}rou, A.-L.},
  \bibinfo{author}{Courtens, E.} \& \bibinfo{author}{Vacher, R.}
\newblock \bibinfo{journal}{\bibinfo{title}{Observation of a doublet in the
  quasielastic central peak of quantum-paraelectric \ch{SrTiO3}}}.
\newblock {\emph{\JournalTitle{Physical Review Letters}}}
  \textbf{\bibinfo{volume}{75}}, \bibinfo{pages}{2416} (\bibinfo{year}{1995}).

\bibitem{koreeda2007second}
\bibinfo{author}{Koreeda, A.}, \bibinfo{author}{Takano, R.} \&
  \bibinfo{author}{Saikan, S.}
\newblock \bibinfo{journal}{\bibinfo{title}{Second sound in \ch{SrTiO3}}}.
\newblock {\emph{\JournalTitle{Phys. Rev. Lett.}}}
  \textbf{\bibinfo{volume}{99}}, \bibinfo{pages}{265502}
  (\bibinfo{year}{2007}).

\bibitem{smontara1996phonon}
\bibinfo{author}{Smontara, A.}, \bibinfo{author}{Lasjaunias, J.} \&
  \bibinfo{author}{Maynard, R.}
\newblock \bibinfo{journal}{\bibinfo{title}{Phonon poiseuille flow in
  quasi-one-dimensional single crystals}}.
\newblock {\emph{\JournalTitle{Physical Review Letters}}}
  \textbf{\bibinfo{volume}{77}}, \bibinfo{pages}{5397} (\bibinfo{year}{1996}).

\bibitem{machida2018observation}
\bibinfo{author}{Machida, Y.} \emph{et~al.}
\newblock \bibinfo{journal}{\bibinfo{title}{Observation of poiseuille flow of
  phonons in black phosphorus}}.
\newblock {\emph{\JournalTitle{Science Advances}}}
  \textbf{\bibinfo{volume}{4}}, \bibinfo{pages}{eaat3374}
  (\bibinfo{year}{2018}).

\bibitem{ding2022observation}
\bibinfo{author}{Ding, Z.} \emph{et~al.}
\newblock \bibinfo{journal}{\bibinfo{title}{Observation of second sound in
  graphite over 200 k}}.
\newblock {\emph{\JournalTitle{Nature Communications}}}
  \textbf{\bibinfo{volume}{13}}, \bibinfo{pages}{1--9} (\bibinfo{year}{2022}).

\bibitem{guo2017heat}
\bibinfo{author}{Guo, Y.} \& \bibinfo{author}{Wang, M.}
\newblock \bibinfo{journal}{\bibinfo{title}{Heat transport in two-dimensional
  materials by directly solving the phonon boltzmann equation under callaway's
  dual relaxation model}}.
\newblock {\emph{\JournalTitle{Physical Review B}}}
  \textbf{\bibinfo{volume}{96}}, \bibinfo{pages}{134312}
  (\bibinfo{year}{2017}).

\bibitem{guo2021size}
\bibinfo{author}{Guo, Y.} \emph{et~al.}
\newblock \bibinfo{journal}{\bibinfo{title}{Size effect on phonon hydrodynamics
  in graphite microstructures and nanostructures}}.
\newblock {\emph{\JournalTitle{Physical Review B}}}
  \textbf{\bibinfo{volume}{104}}, \bibinfo{pages}{075450}
  (\bibinfo{year}{2021}).

\bibitem{machida2020phonon}
\bibinfo{author}{Machida, Y.}, \bibinfo{author}{Matsumoto, N.},
  \bibinfo{author}{Isono, T.} \& \bibinfo{author}{Behnia, K.}
\newblock \bibinfo{journal}{\bibinfo{title}{Phonon hydrodynamics and
  ultrahigh--room-temperature thermal conductivity in thin graphite}}.
\newblock {\emph{\JournalTitle{Science}}} \textbf{\bibinfo{volume}{367}},
  \bibinfo{pages}{309--312} (\bibinfo{year}{2020}).

\bibitem{taniguchi2001spontaneous}
\bibinfo{author}{Taniguchi, T.} \& \bibinfo{author}{Yamaoka, S.}
\newblock \bibinfo{journal}{\bibinfo{title}{Spontaneous nucleation of cubic
  boron nitride single crystal by temperature gradient method under high
  pressure}}.
\newblock {\emph{\JournalTitle{Journal of Crystal Growth}}}
  \textbf{\bibinfo{volume}{222}}, \bibinfo{pages}{549--557}
  (\bibinfo{year}{2001}).

\bibitem{taniguchi2007synthesis}
\bibinfo{author}{Taniguchi, T.} \& \bibinfo{author}{Watanabe, K.}
\newblock \bibinfo{journal}{\bibinfo{title}{Synthesis of high-purity boron
  nitride single crystals under high pressure by using \ch{Ba}--\ch{BN}
  solvent}}.
\newblock {\emph{\JournalTitle{Journal of Crystal Growth}}}
  \textbf{\bibinfo{volume}{303}}, \bibinfo{pages}{525--529}
  (\bibinfo{year}{2007}).

\bibitem{reich2004raman}
\bibinfo{author}{Reich, S.} \& \bibinfo{author}{Thomsen, C.}
\newblock \bibinfo{journal}{\bibinfo{title}{Raman spectroscopy of graphite}}.
\newblock {\emph{\JournalTitle{Philosophical Transactions of the Royal Society
  of London. Series A: Mathematical, Physical and Engineering Sciences}}}
  \textbf{\bibinfo{volume}{362}}, \bibinfo{pages}{2271--2288}
  (\bibinfo{year}{2004}).

\bibitem{ferrari2006raman}
\bibinfo{author}{Ferrari, A.~C.} \emph{et~al.}
\newblock \bibinfo{journal}{\bibinfo{title}{Raman spectrum of graphene and
  graphene layers}}.
\newblock {\emph{\JournalTitle{Physical Review Letters}}}
  \textbf{\bibinfo{volume}{97}}, \bibinfo{pages}{187401}
  (\bibinfo{year}{2006}).

\bibitem{eckmann2012probing}
\bibinfo{author}{Eckmann, A.} \emph{et~al.}
\newblock \bibinfo{journal}{\bibinfo{title}{Probing the nature of defects in
  graphene by raman spectroscopy}}.
\newblock {\emph{\JournalTitle{Nano Letters}}} \textbf{\bibinfo{volume}{12}},
  \bibinfo{pages}{3925--3930} (\bibinfo{year}{2012}).

\bibitem{bae2013ballistic}
\bibinfo{author}{Bae, M.-H.} \emph{et~al.}
\newblock \bibinfo{journal}{\bibinfo{title}{Ballistic to diffusive crossover of
  heat flow in graphene ribbons}}.
\newblock {\emph{\JournalTitle{Nature Communications}}}
  \textbf{\bibinfo{volume}{4}}, \bibinfo{pages}{1--7} (\bibinfo{year}{2013}).

\bibitem{xu2014length}
\bibinfo{author}{Xu, X.} \emph{et~al.}
\newblock \bibinfo{journal}{\bibinfo{title}{Length-dependent thermal
  conductivity in suspended single-layer graphene}}.
\newblock {\emph{\JournalTitle{Nature Communications}}}
  \textbf{\bibinfo{volume}{5}}, \bibinfo{pages}{1--6} (\bibinfo{year}{2014}).

\bibitem{fugallo2014thermal}
\bibinfo{author}{Fugallo, G.} \emph{et~al.}
\newblock \bibinfo{journal}{\bibinfo{title}{Thermal conductivity of graphene
  and graphite: collective excitations and mean free paths}}.
\newblock {\emph{\JournalTitle{Nano Letters}}} \textbf{\bibinfo{volume}{14}},
  \bibinfo{pages}{6109--6114} (\bibinfo{year}{2014}).

\bibitem{inyushkin2004isotope}
\bibinfo{author}{Inyushkin, A.}, \bibinfo{author}{Taldenkov, A.},
  \bibinfo{author}{Gibin, A.}, \bibinfo{author}{Gusev, A.} \&
  \bibinfo{author}{Pohl, H.-J.}
\newblock \bibinfo{journal}{\bibinfo{title}{On the isotope effect in thermal
  conductivity of silicon}}.
\newblock {\emph{\JournalTitle{Physica Status Solidi (C)}}}
  \textbf{\bibinfo{volume}{1}}, \bibinfo{pages}{2995--2998}
  (\bibinfo{year}{2004}).

\bibitem{zheng2019thermal}
\bibinfo{author}{Zheng, Q.} \emph{et~al.}
\newblock \bibinfo{journal}{\bibinfo{title}{Thermal conductivity of
  $\mathrm{GaN}$, $^{71}\mathrm{GaN}$, and $\mathrm{SiC}$ from 150
  \uppercase{K} to 850 \uppercase{K}}}.
\newblock {\emph{\JournalTitle{Physical Review Materials}}}
  \textbf{\bibinfo{volume}{3}}, \bibinfo{pages}{014601} (\bibinfo{year}{2019}).

\bibitem{zheng2018high}
\bibinfo{author}{Zheng, Q.} \emph{et~al.}
\newblock \bibinfo{journal}{\bibinfo{title}{High thermal conductivity in
  isotopically enriched cubic boron phosphide}}.
\newblock {\emph{\JournalTitle{Advanced Functional Materials}}}
  \textbf{\bibinfo{volume}{28}}, \bibinfo{pages}{1805116}
  (\bibinfo{year}{2018}).

\bibitem{chen2012thermal}
\bibinfo{author}{Chen, S.} \emph{et~al.}
\newblock \bibinfo{journal}{\bibinfo{title}{Thermal conductivity of
  isotopically modified graphene}}.
\newblock {\emph{\JournalTitle{Nature Materials}}}
  \textbf{\bibinfo{volume}{11}}, \bibinfo{pages}{203--207}
  (\bibinfo{year}{2012}).

\bibitem{anthony1990thermal}
\bibinfo{author}{Anthony, T.} \emph{et~al.}
\newblock \bibinfo{journal}{\bibinfo{title}{Thermal diffusivity of isotopically
  enriched $^{12}\mathrm{C}$ diamond}}.
\newblock {\emph{\JournalTitle{Physical Review B}}}
  \textbf{\bibinfo{volume}{42}}, \bibinfo{pages}{1104} (\bibinfo{year}{1990}).

\bibitem{gurzhi1964thermal}
\bibinfo{author}{Gurzhi, R.}
\newblock \bibinfo{journal}{\bibinfo{title}{Thermal conductivity of dielectrics
  and ferrodielectrics at low temperatures}}.
\newblock {\emph{\JournalTitle{Sov. Phys. JETP}}}
  \textbf{\bibinfo{volume}{19}}, \bibinfo{pages}{490} (\bibinfo{year}{1964}).

\bibitem{gurzhi1968hydrodynamic}
\bibinfo{author}{Gurzhi, R.}
\newblock \bibinfo{journal}{\bibinfo{title}{Hydrodynamic effects in solids at
  low temperature}}.
\newblock {\emph{\JournalTitle{Soviet Physics Uspekhi}}}
  \textbf{\bibinfo{volume}{11}}, \bibinfo{pages}{255} (\bibinfo{year}{1968}).

\bibitem{kopylov1974investigation}
\bibinfo{author}{Kopylov, V.} \& \bibinfo{author}{Mezhov-Deglin, L.}
\newblock \bibinfo{journal}{\bibinfo{title}{Investigation of the kinetic
  coefficients of bismuth at helium temperatures}}.
\newblock {\emph{\JournalTitle{Soviet Journal of Experimental and Theoretical
  Physics}}} \textbf{\bibinfo{volume}{38}}, \bibinfo{pages}{357}
  (\bibinfo{year}{1974}).

\bibitem{mezhov1966measurement}
\bibinfo{author}{Mezhov-Deglin, L.}
\newblock \bibinfo{journal}{\bibinfo{title}{Measurement of the thermal
  conductivity of crystalline \ch{He4}}}.
\newblock {\emph{\JournalTitle{Sov. Phys. JETP}}}
  \textbf{\bibinfo{volume}{22}}, \bibinfo{pages}{47} (\bibinfo{year}{1966}).

\bibitem{alexander1980low}
\bibinfo{author}{Alexander, M.~G.}, \bibinfo{author}{Goshorn, D.~P.} \&
  \bibinfo{author}{Onn, D.~G.}
\newblock \bibinfo{journal}{\bibinfo{title}{{Low-temperature specific heat of
  the graphite intercalation compounds K${\mathrm{C}}_{8}$,
  Cs${\mathrm{C}}_{8}$, Rb${\mathrm{C}}_{8}$, and their parent highly oriented
  pyrolytic graphite}}}.
\newblock {\emph{\JournalTitle{Physical Review B}}}
  \textbf{\bibinfo{volume}{22}}, \bibinfo{pages}{4535} (\bibinfo{year}{1980}).

\bibitem{guyer1966thermal}
\bibinfo{author}{Guyer, R.} \& \bibinfo{author}{Krumhansl, J.}
\newblock \bibinfo{journal}{\bibinfo{title}{Thermal conductivity, second sound,
  and phonon hydrodynamic phenomena in nonmetallic crystals}}.
\newblock {\emph{\JournalTitle{Physical Review}}}
  \textbf{\bibinfo{volume}{148}}, \bibinfo{pages}{778} (\bibinfo{year}{1966}).

\bibitem{chen2005}
\bibinfo{author}{Chen, G.}
\newblock \emph{\bibinfo{title}{Nanoscale Energy Transport and Conversion}}
  (\bibinfo{publisher}{Oxford University Press Inc}, \bibinfo{address}{New
  York}, \bibinfo{year}{2005}).

\bibitem{kittel2004}
\bibinfo{author}{Kittel, C.}
\newblock \emph{\bibinfo{title}{Introduction to Solid State Physics}}
  (\bibinfo{publisher}{John Wiley and Sons, Inc}, \bibinfo{address}{New York},
  \bibinfo{year}{2004}).

\bibitem{alofi2014theory}
\bibinfo{author}{Alofi, A. S.~S.}
\newblock \emph{\bibinfo{title}{Theory of Phonon Thermal Transport in Graphene
  and Graphite}}.
\newblock Ph.D. thesis, \bibinfo{school}{University of Exeter}
  (\bibinfo{year}{2014}).

\bibitem{li2019crossover}
\bibinfo{author}{Li, X.} \& \bibinfo{author}{Lee, S.}
\newblock \bibinfo{journal}{\bibinfo{title}{Crossover of ballistic,
  hydrodynamic, and diffusive phonon transport in suspended graphene}}.
\newblock {\emph{\JournalTitle{Physical Review B}}}
  \textbf{\bibinfo{volume}{99}}, \bibinfo{pages}{085202}
  (\bibinfo{year}{2019}).

\bibitem{bausch1972thermal}
\bibinfo{author}{Bausch, W.}
\newblock \bibinfo{journal}{\bibinfo{title}{Thermal conductivity and poiseuille
  flow of phonons in dielectric films and plates}}.
\newblock {\emph{\JournalTitle{Physica Status Solidi (B)}}}
  \textbf{\bibinfo{volume}{52}}, \bibinfo{pages}{253--262}
  (\bibinfo{year}{1972}).

\bibitem{maire2014reduced}
\bibinfo{author}{Maire, J.} \& \bibinfo{author}{Nomura, M.}
\newblock \bibinfo{journal}{\bibinfo{title}{Reduced thermal conductivities of
  si one-dimensional periodic structure and nanowire}}.
\newblock {\emph{\JournalTitle{Japanese Journal of Applied Physics}}}
  \textbf{\bibinfo{volume}{53}}, \bibinfo{pages}{06JE09}
  (\bibinfo{year}{2014}).

\bibitem{pope2001description}
\bibinfo{author}{Pope, A.}, \bibinfo{author}{Zawilski, B.} \&
  \bibinfo{author}{Tritt, T.}
\newblock \bibinfo{journal}{\bibinfo{title}{Description of removable sample
  mount apparatus for rapid thermal conductivity measurements}}.
\newblock {\emph{\JournalTitle{Cryogenics}}} \textbf{\bibinfo{volume}{41}},
  \bibinfo{pages}{725--731} (\bibinfo{year}{2001}).

\bibitem{maire2015thermal}
\bibinfo{author}{Maire, J.}
\newblock \emph{\bibinfo{title}{Thermal phonon transport in silicon
  nanosturctures}}.
\newblock Ph.D. thesis, \bibinfo{school}{The University of Tokyo}
  (\bibinfo{year}{2015}).

\bibitem{nihira2003temperature}
\bibinfo{author}{Nihira, T.} \& \bibinfo{author}{Iwata, T.}
\newblock \bibinfo{journal}{\bibinfo{title}{Temperature dependence of lattice
  vibrations and analysis of the specific heat of graphite}}.
\newblock {\emph{\JournalTitle{Physical Review B}}}
  \textbf{\bibinfo{volume}{68}}, \bibinfo{pages}{134305}
  (\bibinfo{year}{2003}).

\bibitem{ho1972thermal}
\bibinfo{author}{Ho, C.~Y.}, \bibinfo{author}{Powell, R.~W.} \&
  \bibinfo{author}{Liley, P.~E.}
\newblock \bibinfo{journal}{\bibinfo{title}{Thermal conductivity of the
  elements}}.
\newblock {\emph{\JournalTitle{Journal of Physical and Chemical Reference
  Data}}} \textbf{\bibinfo{volume}{1}}, \bibinfo{pages}{279--421}
  (\bibinfo{year}{1972}).

\bibitem{li2014shengbte}
\bibinfo{author}{Li, W.}, \bibinfo{author}{Carrete, J.},
  \bibinfo{author}{Katcho, N.~A.} \& \bibinfo{author}{Mingo, N.}
\newblock \bibinfo{journal}{\bibinfo{title}{Shengbte: A solver of the boltzmann
  transport equation for phonons}}.
\newblock {\emph{\JournalTitle{Computer Physics Communications}}}
  \textbf{\bibinfo{volume}{185}}, \bibinfo{pages}{1747--1758}
  (\bibinfo{year}{2014}).

\bibitem{giannozzi2009quantum}
\bibinfo{author}{Giannozzi, P.} \emph{et~al.}
\newblock \bibinfo{journal}{\bibinfo{title}{Quantum espresso: a modular and
  open-source software project for quantum simulations of materials}}.
\newblock {\emph{\JournalTitle{Journal of Physics Condensed Matter}}}
  \textbf{\bibinfo{volume}{21}}, \bibinfo{pages}{395502}
  (\bibinfo{year}{2009}).

\bibitem{lindsay2011flexural}
\bibinfo{author}{Lindsay, L.}, \bibinfo{author}{Broido, D.} \&
  \bibinfo{author}{Mingo, N.}
\newblock \bibinfo{journal}{\bibinfo{title}{Flexural phonons and thermal
  transport in multilayer graphene and graphite}}.
\newblock {\emph{\JournalTitle{Physical Review B}}}
  \textbf{\bibinfo{volume}{83}}, \bibinfo{pages}{235428}
  (\bibinfo{year}{2011}).

\bibitem{ghosh2010dimensional}
\bibinfo{author}{Ghosh, S.} \emph{et~al.}
\newblock \bibinfo{journal}{\bibinfo{title}{Dimensional crossover of thermal
  transport in few-layer graphene}}.
\newblock {\emph{\JournalTitle{Nature Materials}}}
  \textbf{\bibinfo{volume}{9}}, \bibinfo{pages}{555--558}
  (\bibinfo{year}{2010}).

\end{thebibliography}


\section*{Acknowledgements} 
This work was supported by JSPS KAKENHI (Grant Numbers 21H04635, JP19H01820, JP20H00127, JP21H05232, JP21H05233, and 21H05234) the Postdoctoral Fellowship of JSPS (P19353), and CREST JST (Grant Numbers JPMJCR19Q3 and JPMJCR20B4).

\section*{Author contributions}

X.H., Y.G. and M.N. conceived this study. X.H. and Y.W. designed and conducted the fabrication. X.H. performed the TDTR and Raman spectra measurements, analyzed the results, and wrote the manuscript. S.M. and T.M. provided the graphite samples, mechanical exfoliation technique and Raman spectroscopy technique. S.M. contributed to designing the fabrication and Raman spectra measurements. K.W. and T.T. synthesized the isotopically-purified graphite crystals. Y.G. developed the theoretical model and contributed to writing the manuscript. Z.Z. and S.V. contributed to interpreting the results. M.N. supervised this work. All authors contributed to discussing the results and manuscript revision.

\section*{Competing interests}
The authors declare that they have no competing interests.

\newpage

\section*{\LARGE Supporting Information}
\vspace{2cm}
\section*{Supplementary Note 1. Sample characterization}

Using time-of-flight secondary ion mass spectrometry (TOF-SIMS), we first measured the $^{13}$C concentration in our isotopically-purified graphite crystals. As shown in Supplementary Fig.~\ref{lab_figs1}a, a value of 0.02\% was obtained and proved the isotopic purity of the crystals. As our reference sample, we employed natural graphite ("Flaggy flakes") with a naturally occurring $^{13}$C abundance of 1.1\%, which is the cleanest commercial sample for preparing graphene with large areas. 

In addition, we performed Raman spectroscopy to characterize the identical crystal quality of isotopically-purified and natural graphite samples. In Supplementary Fig.~\ref{lab_figs1}b, the Raman spectra show the same Raman selected G and 2D peaks for both samples. We observed no other initial Raman peaks caused by defects which would destroy the symmetry of carbon hexagonal lattice, and turned out the perfect crystalline for both two graphite samples \cite{reich2004raman,ferrari2006raman,eckmann2012probing}. All the aforementioned sample characterizations confirmed that the only difference between the two samples comes from the $^{13}$C isotope concentration, and justified the investigation of in-plane thermal conductivity and phonon Poiseuille flow on both isotopically-purified and natural graphite samples in this work.

\section*{Supplementary Note 2. Examining the criterion of phonon Poiseuille flow in graphite}


To have a more quantitative understanding of the satisfaction of hydrodynamic window condition ($l_{N} \ll W, ~l_{R}l_{N}\gg W^{2}$) in our purified graphite ribbon, we show the MFPs of normal and resistive scatterings of bending acoustic (BA) phonons (at $k_{z}=0$) in natural and isotopically-purified graphite at 60~K obtained by our first-principles modeling in Supplementary Fig.~\ref{lab_figs6}. Note that the resistive scattering here is basically presented by the isotope scattering since the Umklapp process is rare at lower temperatures. In both purified (0.02\% $^{13}$C) and natural (1.1\% $^{13}$C) cases, the sample widths are much larger than the MFP of normal scattering ($l_{N} \ll W$). In the natural graphite ribbon of the present study, the MFP of isotope scattering is around one order of magnitude larger than the sample width (5~\textmu m), such that $l_{R}l_{N}\sim W^{2}$. In other words, the graphite ribbon with a natural abundance of $^{13}$C does not satisfy the window condition, which is instead valid in the purified ribbon since the MFP of isotope scattering is around two orders of magnitude larger than the sample width ($l_{R}l_{N}\gg W^{2}$). This justifies the present observation of phonon Poiseuille flow in the isotopically-purified graphite ribbon with a width of 5~\textmu m while not in the natural graphite ribbon with the same width. 

The calculation of the ballistic thermal conductance relies on the used atomic interaction potential. For a comparison, we also calculate the temperature-dependent heat capacity and ballistic thermal conductance of graphite based on an empirical potential \cite{lindsay2011flexural}. The optimized Tersoff potential and Lennard-Jones potential are adopted for the in-plane and inter-layer interactions respectively. As shown in Supplementary Fig.~\ref{lab_figs4}(b), although there is some difference between the absolute values at low temperature, the temperature scaling behaviors of ballistic thermal conductance are almost the same by the first-principles (DFT) calculation and by the empirical potential. Therefore, the theoretical model shall have minor influence on the conclusion of our work. Still we recommend to adopt the first-principles method to calculate the ballistic thermal conductance as the obtained phonon dispersion better reproduces the experimental data of heat capacity, as seen in Supplementary Fig.~\ref{lab_figs4}(a).


\section*{Supplementary Note 3. Comparison of window conditions for phonon Poiseuille flow and second sound in graphite}

A commonly admitted condition of the second sound is that the excitation pulse frequency is smaller than the normal scattering rate but larger than the resistive scattering rate ($\tau_{N}^{-1} \gg \Omega \gg \tau_{R}^{-1}$) \cite{lee2015hydrodynamic}, or is equivalently the dominance of normal scattering over the resistive scattering ($l_{N} \ll l_{ex} \ll l_{R}$) \cite{guyer1966thermal}, with $l_{ex}$ referring to the length of the external excitation. The observation of second sound has been recently reported in the HOPG sample with natural isotope at 100~K \cite{huberman2019observation}, followed by a very recent update at 200~K \cite{ding2022observation} using an improved version of the same technique. In these two studies, the transient thermal grating (TTG) method was used to generate a periodically oscillating temperature field and the decay of the temperature amplitude was measured to indicate the second sound. However, observing the steady-state hydrodynamic phenomenon, namely, the phonon Poiseuille flow, is expected to be more challenging than in the second sound case. As proposed by Guyer \textit{et al.}, phonon Poiseuille flow appears only under the following conditions\cite{guyer1966thermal}: $l_{N} \ll W, ~l_{R}l_{N}\gg W^{2}$. Again, we adopt Supplementary Fig.~\ref{lab_figs6} for a more quantitative understanding of the isotope effect on observing phonon Poiseuille flow and second sound. As demonstrated in the main text and explained in Supplementary Note 2, the phonon Poiseuille flow is only observed in the isotopically-purified graphite ribbon with a width of 5~\textmu m while not in the natural graphite ribbon with the same width. For the case of the second sound, the MFP of isotope scattering in the isotopically-purified sample and natural one is around three and two orders of magnitude larger than the MFP of normal scattering respectively, as seen in Supplementary Fig.~\ref{lab_figs6}. Hence, the window condition to observe the second sound via TTG with a grating period in-between the MFPs of isotope scattering and normal scattering is satisfied in both purified and natural cases.

\section*{Supplementary Note 4. Thermal boundary conductance between graphite and metals}

As an unavoidable issue in TDTR measurement, the influence of thermal boundary conductance (TBC) between the aluminum transducer and graphite is also investigated using the finite element method (FEM). We first built a heat dissipation model in COMSOL Multiphysics with the same structures as our \textmu-TDTR measurement. In the simulation, we applied a Gaussian heat flux pulse to simulate the pump laser focused on the aluminum transducer in the experiment, as shown in Supplementary Figs.~\ref{lab_figs9}a,b. Then, by varying the TBC of aluminum/graphite interface, we studied its effects on the decay curve and observed a negligible change of decay curves in Supplementary Fig.~\ref{lab_figs9}c. To explain, we checked the temperature difference between aluminum and graphite layers on the real-time scale after heat flux injection. We found that the aluminum pad and graphite island reached thermal equilibrium within 1\textmu s (Supplementary Fig.~\ref{lab_figs9}d) while heat conducts in much longer time-scale through the ribbon. Hence, we conclude a negligible impact of TBC between aluminum and graphite layers on the results in our microsecond-scale experiments. 


\newpage

\makeatletter
\renewcommand{\fnum@figure}{Supplementary Fig. \thefigure}
\setcounter{figure}{0}
\makeatother

\begin{figure}[H]
\centering
\includegraphics[width=1\textwidth]{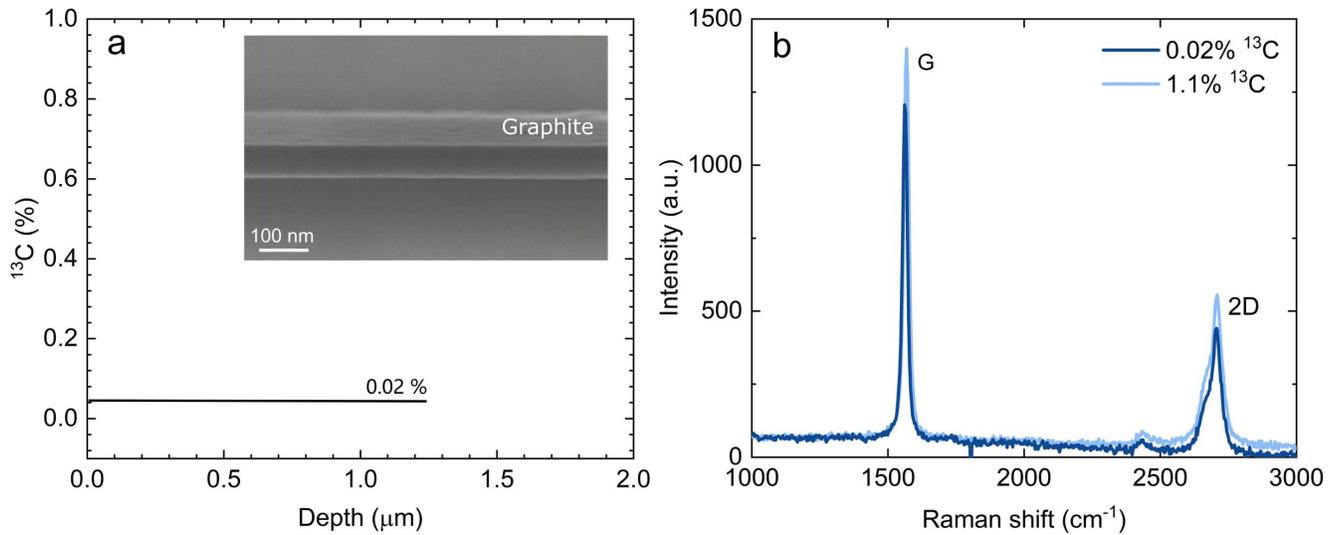}
\caption{\textbf{Sample characterization using TOF-SIMS and Raman spectroscopy.} \textbf{a} $^{13}$C concentration measured using time-of-flight secondary ion mass spectrometry (TOF-SIMS). Inset: cross-section view of scanning electron microscope (SEM) image of the suspended isotopically-purified graphite ribbon. It indicates the thickness of which is approximately 65~nm. Note that the thickness of natural ribbon is 90~nm, and we assume a negligible thickness-dependence due to the weak van der Waals interaction along the out-of-plane direction as justified in previous works \cite{lindsay2011flexural,ghosh2010dimensional}. \textbf{b} Raman spectra of both isotopically-purified (0.02\% $^{13}$C) and natural (1.1\% $^{13}$C) graphite samples.}
\label{lab_figs1}
\end{figure}

\begin{figure}[H]
\centering
\includegraphics[width=1\textwidth]{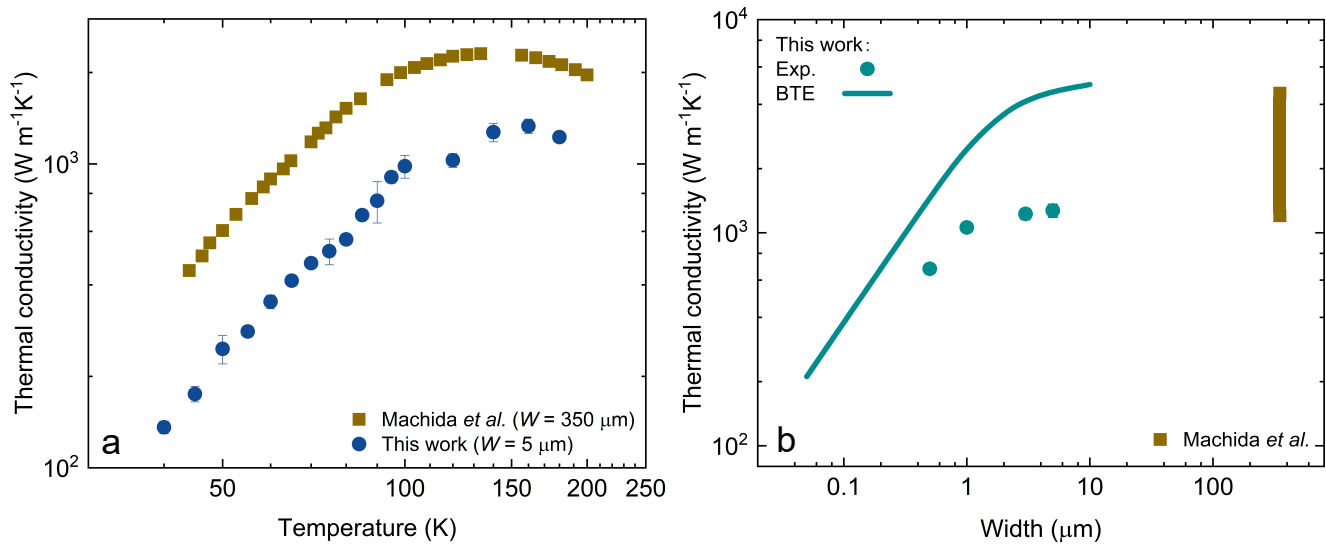}
\caption{\textbf{Comparison of in-plane thermal conductivity of submicroscale graphite ribbons with bulk HOPG.} \textbf{a} Temperature-dependent in-plane thermal conductivity of our isotopically-purified graphite ribbon (Length: 30~\textmu m, Width: 5~\textmu m, Thickness: 65~nm) and bulk HOPG (Length: 6500~\textmu m, Width: 350~\textmu m, Thickness: 240~\textmu m) \cite{machida2020phonon}. \textbf{b} Width-dependent in-plane thermal conductivity of our isotopically-purified graphite ribbon (Length: 30~\textmu m, Thickness: 65~nm) and bulk HOPG (Length: 6500~\textmu m, Width: 350~\textmu m, Thickness: 8.5$-$580~\textmu m) \cite{machida2020phonon} at 150~K. The absolute value of our experimental data is lower than that of our modelling results by BTE with first-principles inputs, which might be caused by the additional resistive scattering of phonons induced by unknown defect or contamination in sample fabrication. However, our experimental width-dependence of thermal conductivity shows good qualitative consistency with our calculated results. Moreover, both experimental and calculated results of $\kappa$/$G_\textup{ballistic}$ are normalized by their values at 10~K (ballistic limit), which show very consistent trends to demonstrate the hydrodynamic phonon transport in this work.}
\label{lab_figs2}
\end{figure}

\begin{figure}[H]
\centering
\includegraphics[width=1\textwidth]{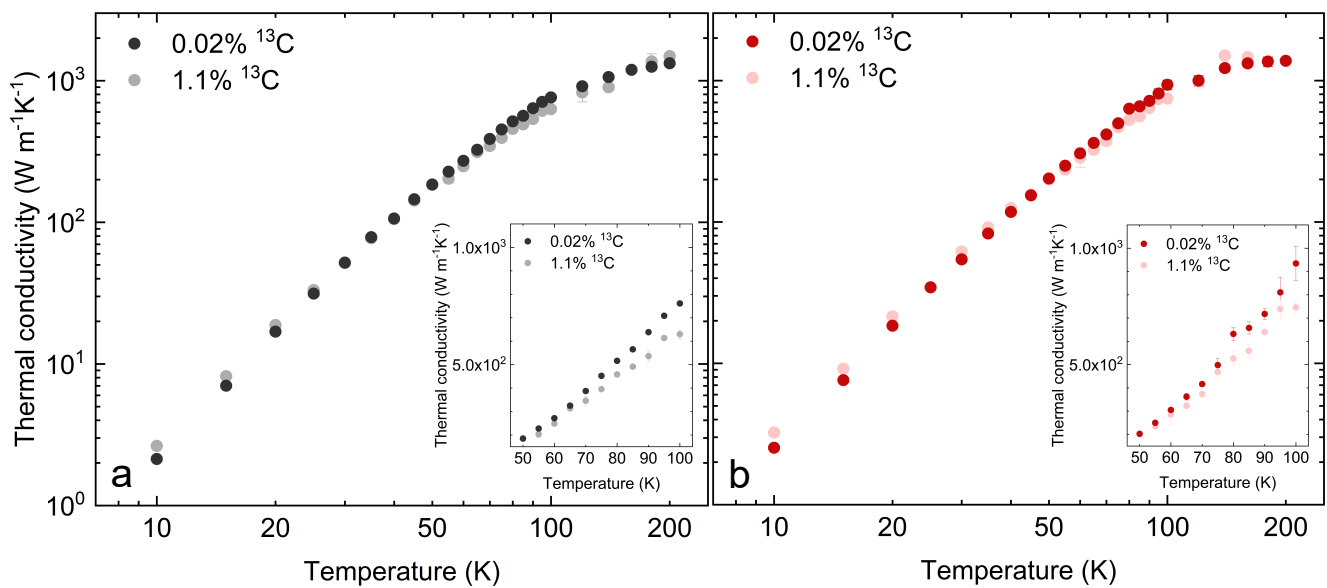}
\caption{\textbf{Impact of isotope on the thermal conductivity of graphite ribbons.} \textbf{a,b} In-plane thermal conductivity as a function of temperature for both isotopically-purified (0.02\% $^{13}$C) and natural (1.1\% $^{13}$C) graphite with the width of (\textbf{a}) 1~\textmu m and (\textbf{b}) 3~\textmu m. As temperature cools down to 100~K, the weakening of Umklapp scattering causes the dominant isotopic effect, which attributes to the thermal conductivity reduction in natural graphite ribbons (gray and pink dots) compared to that in isotopically-purified ribbons (black and red dots). Note that the actual widths of these two natural graphite ribbons are 500 nm wider than that of the isotopically-purified ones due to the deviation in fabrication, resulting in the minor flip of thermal conductivities at very low temperatures. Insets: thermal conductivity data in linear scale from 50 to 100~K.}
\label{lab_figs3}
\end{figure}

\begin{figure}[H]
\centering
\includegraphics[width=1.0\textwidth]{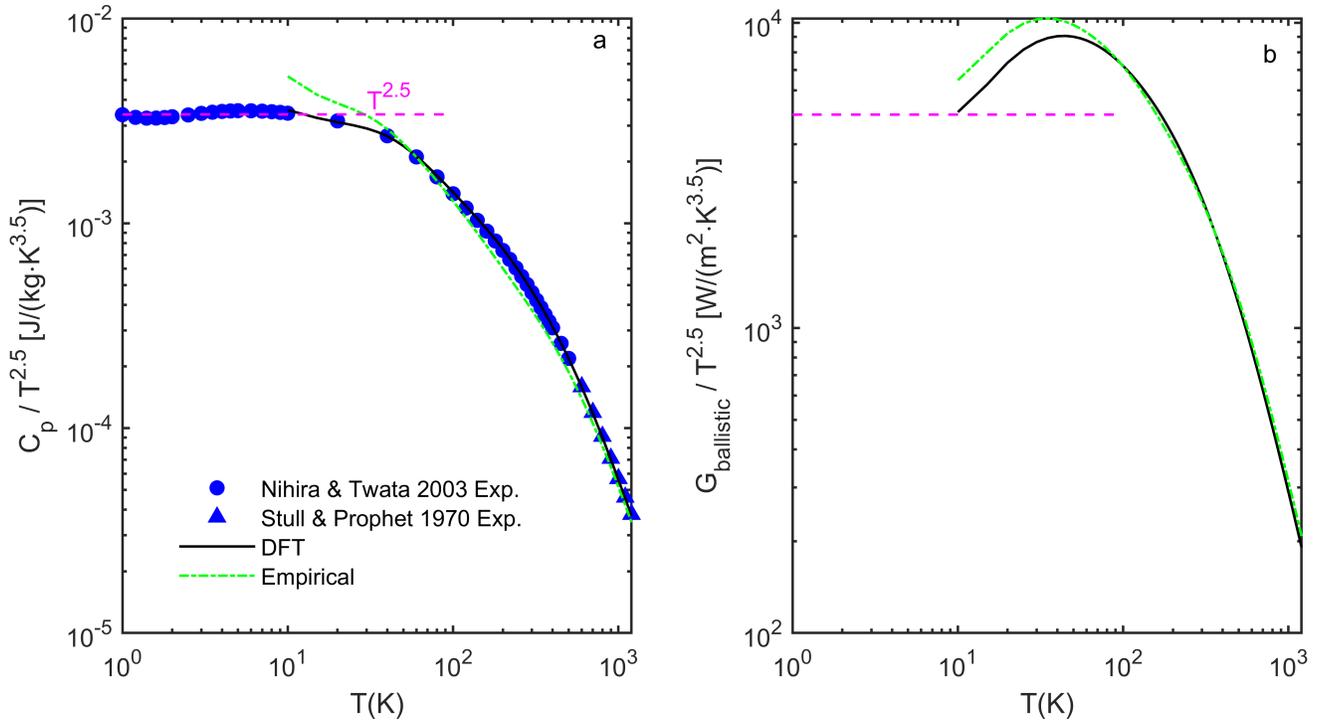}
\caption{Temperature scalings of (\textbf{a}) heat capacity and (\textbf{b}) ballistic thermal conductance of graphite. The discrete symbols represent the experimental data, whereas the solid line and dash-dotted line represents the calculation results by DFT (density functional theory) and by empirical atomic interaction potential respectively. The dashed line denotes the low-temperature limit of $C_{p} \sim T^{2.5}$ and $G_{ballistic} \sim T^{2.5}$. Both theoretical calculations are down to only 10~K due to insufficient resolution of the first Brillouin zone around $\Gamma$ point below 10~K.}
\label{lab_figs4}
\end{figure}

\begin{figure}[H]
\centering
\includegraphics[width=1\textwidth]{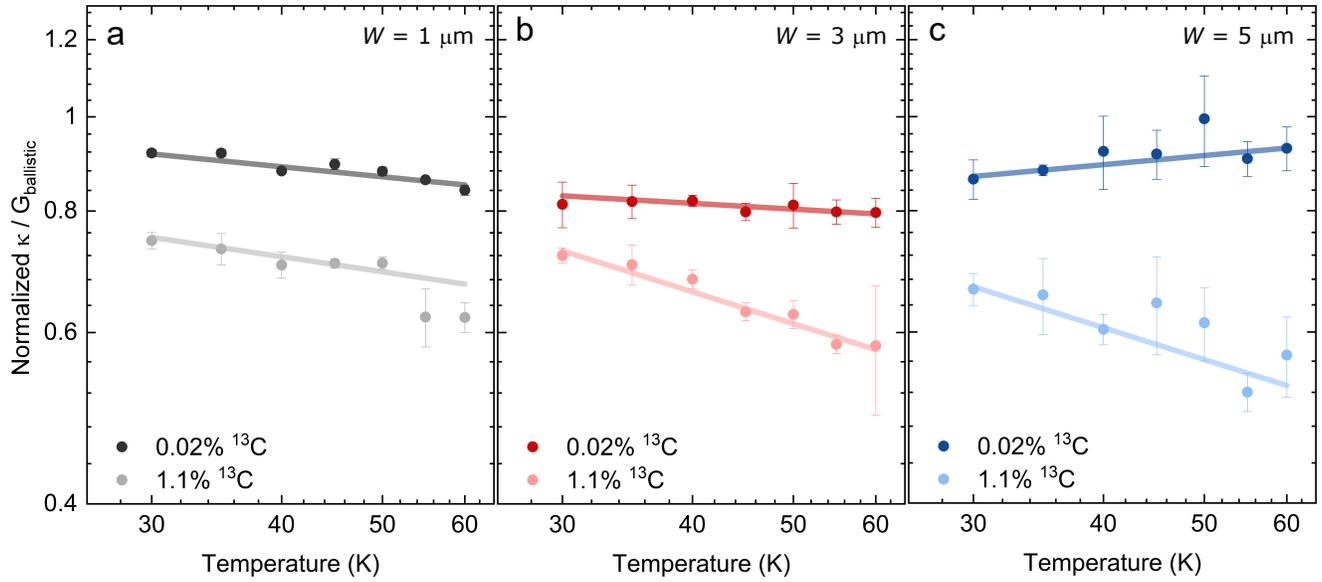}
\caption{\textbf{Transition from ballistic to hydrodynamic heat transport and the isotope impact on phonon Poiseuille flow.} Thermal conductivity over ballistic thermal conductance ($\kappa$/$G_\textup{ballistic}$) normalised by its value at 10~K as a function of temperature from 30 to 60~K for the graphite ribbons with the width of \textbf{(a)} 1~\textmu m, \textbf{(b)} 3~\textmu m and \textbf{(c)} 5~\textmu m. The black (gray), red (pink), and dark blue (light blue) dots represent the experimental data of isotopically-purified (natural) graphite ribbons. Solid lines show the linear fitting of experimental data. Note that the actual width of the natural graphite ribbons are 500~nm to 1~\textmu m wider than that of the isotopically-purified ones. A more significant difference between the slopes of natural and purified samples with the widening of the ribbon is expected when their widths are exactly the same.}
\label{lab_figs5}
\end{figure}

\begin{figure}[H]
\centering
\includegraphics[width=0.7\textwidth]{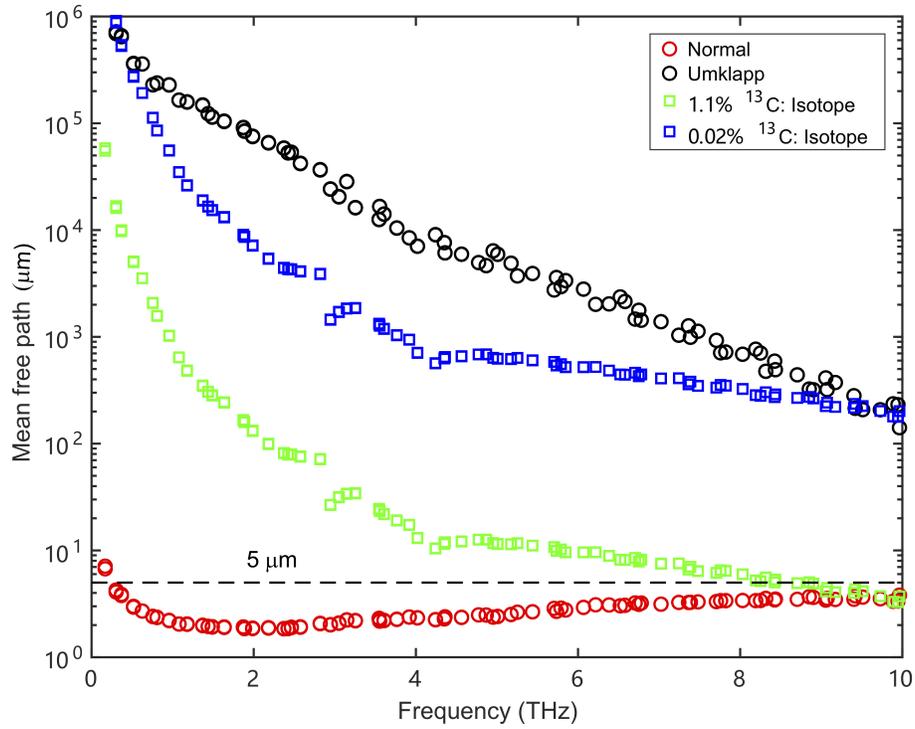}
\caption{\textbf{Mean free paths of different phonon scattering processes in graphite at 60~K.} The red and black circles denote the mean free path of normal and Umklapp scatterings respectively. The green and blue squares denote the mean free path of isotope scattering in natural (1.1\% $^{13}$C) and isotopically-purified (0.02\% $^{13}$C) graphite respectively. The results of the bending acoustic (BA) phonons (at $k_{z}=0$) which dominate the hydrodynamic transport are shown here. The reference size of 5~\textmu m is the sample width of the present graphite ribbon.} 
\label{lab_figs6}
\end{figure}

\begin{figure}[H]
\centering
\includegraphics[width=0.7\textwidth]{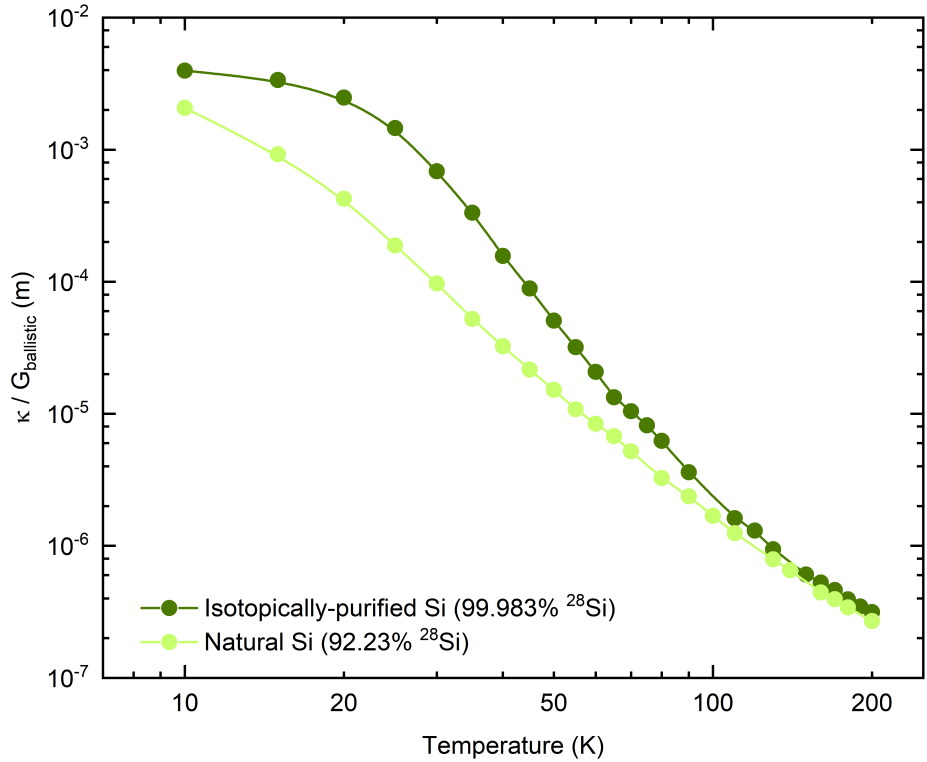}
\caption{\textbf{Examining the criterion of phonon Poiseuille flow in silicon.} The ratio of thermal conductivity ($\kappa$) over ballistic thermal conductance ($G_\textup{ballistic}$) (the present criterion) as a function of temperature for bulk silicon with purified (99.983\%) and natural abundance (92.23\%) $^{28}$Si isotope. The experimental thermal conductivity of silicon is obtained from Ref.\cite{inyushkin2004isotope}. The ballistic thermal conductance for silicon is calculated by the first-principles method.}
\label{lab_figs7}
\end{figure}

\begin{figure}[H]
\centering
\includegraphics[width=1\textwidth]{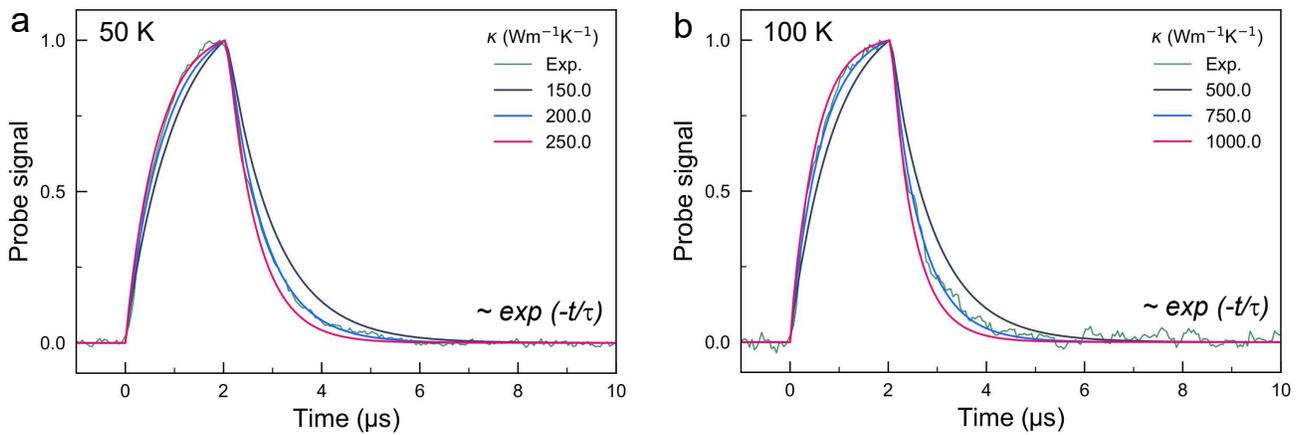}
\caption{\textbf{Fitting experimental data with FEM simulation.} At two typical temperatures of (\textbf{a}) 50~K and (\textbf{b}) 100~K, we demonstrated the fitting of exponential decay curves obtained in TDTR by the finite element method (FEM) simulations to ensure the accurate extraction of thermal conductivity of our graphite ribbon samples.}
\label{lab_figs8}
\end{figure}

\begin{figure}[H]
\centering
\includegraphics[width=1\textwidth]{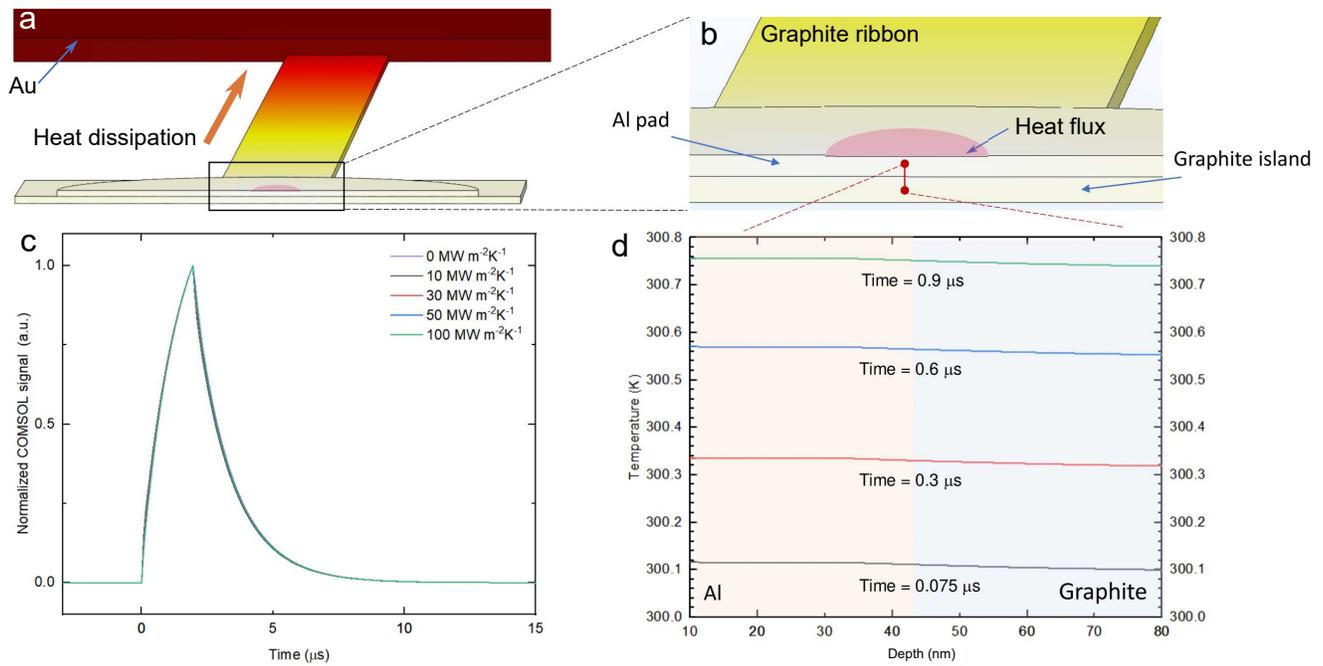}
\caption{\textbf{Investigation of thermal boundary conductance between graphite and metals in FEM simulation.} \textbf{a,b} Illustration of heat dissipation in FEM model. \textbf{c} The dependency of exponential decay on thermal boundary conductance (TBC) between graphite and aluminum. \textbf{d} Real-time evolution of temperature distributions around aluminum/graphite interface.}
\label{lab_figs9}
\end{figure}



\end{document}